\documentclass{article}

\usepackage{dcolumn}
\usepackage[utf8]{inputenc}
\usepackage{fancyvrb}
\usepackage{amsmath,amssymb}
\usepackage{subcaption}
\usepackage{enumitem}
\usepackage{graphicx}
\usepackage{hyperref}
\usepackage{xcolor}
\usepackage{dsfont}
\usepackage[ruled]{algorithm2e}
\usepackage{booktabs}

\newcolumntype{d}[1]{D{.}{.}{#1}}
\newcommand{\SNAS}{\texttt{SL}$_{\texttt{NAS}}$}
\newcommand{\RBF}{\texttt{RBF}}
\newcommand{\GAN}{\texttt{GAN}}

\newcommand{\NN}{\texttt{NN}}
\newcommand{\NAS}{\texttt{NAS}}
\newcommand{\RSS}{\texttt{RSS}}

\newcommand{\RSSI}{\texttt{RSSI}}

\newcommand{\TV}{\texttt{TV}}

\newcommand{\UNet}{\texttt{UNet}}
\newcommand{\LoS}{\texttt{LoS}}
\newcommand{\NLoS}{\texttt{NLoS}}
\newcommand{\GPS}{\texttt{GPS}}

\DeclareMathOperator*{\argmin}{argmin}

\newcommand{\Rep}[1]{\color{black}#1\color{black}}

\title{Deep Learning  with Partially Labeled Data for Radio Map Reconstruction}

\author{Alkesandra Malkova$^\dagger$, Massih-Reza Amini$^\dagger$, Beno\^it Denis$^\ddagger$, Christophe Villien$^\ddagger$\\
$^\dagger$ LIG-APTIKAL, Universit\'e Grenoble Alpes,\\
700 Av. Centrale, 38401 Saint-Martin-d’H\`eres, France\\
\{Firstname.Lastname\}@univ-grenoble-alpes.fr\\
$^\ddagger$ CEA-Leti, 17 Av. des Martyrs, Grenoble, France\\
\{Firstname.Lastname\}@cea.fr
}

\date{}

\begin{document}
\maketitle

\begin{abstract}
In this paper, we address the problem of Received Signal Strength map reconstruction based on location-dependent radio measurements and utilizing side knowledge about the local region; for example, city plan, terrain height, gateway position.
Depending on the quantity of such prior side information, we employ Neural Architecture Search to find an optimized Neural Network model with the best architecture for each of the supposed settings.
We demonstrate that using additional side information enhances the final accuracy of the Received Signal Strength map reconstruction on three datasets that correspond to three major cities, particularly in sub-areas near the gateways where larger variations of the average received signal power are typically observed. 
\end{abstract}

\section{Introduction}
Retrieving the exact position of the connected objects has become an important feature of the Internet of Things (\texttt{IoT}). Such connected objects have indeed been widespread over the last few years thanks to the low cost of the radio integrated chips and sensors and their possibility of being embedded in plurality of the devices.

By this they can help in fast development of large-scale physical monitoring and crowdsensing systems (like smart cities, factories, transportation, etc.). For the location-dependent application and services these abilities to associate accurate location with physical data gives huge opportunities \cite{localization_iot_survey}. For example, the fine-grain and dynamic update of air pollution and/or weather maps could benefit from geo-referenced mobile sensing \cite{SARWS}
(e.g., aboard taxis, buses, bicycles...), thus continuously complementing the data from static stations. One of the localization techniques is Global Positioning System (\texttt{GPS}) which has been widely used over the past decades. More recently, low-cost advanced \texttt{GPS} solutions were proposed (like \texttt{RTK}, Bi-band,..), but still they suffer from high energy consumption which is not suitable for \texttt{IoT} applications. 

As an alternative, one can opportunistically measure location-dependent radio metrics, like Received Signal Strength (Indicator) (\texttt{RSSI}), Time (Difference) of Arrival, Angle of Arrival, etc., because these sensor nodes communicate with one/several gateways at the same time (e.g. while sending a packet to the infrastructure to store this data in the cloud/server). Based on these metrics, there exist several methods to determine the node position: trilateration, triangulation, proximity detection, or fingerprinting \cite{Yu09, book_wireless_sensor_networks, Cheng12}. \Rep{We will focus on the last approach -- fingerprinting \cite{ LoRa-fingerprint_outdoor}-- which requires (ideally) full map of mentioned above radio metrics, covering the zone of interest. However, collecting metrics in each point of the zone of interest is impractical and time costly in real-world scenarios, therefore most approaches rely on sparse and non-uniformly distributed measurements. 

In this sense, classical map interpolation techniques such as \RBF{} \cite{rbf_interp, powell1987_rbf, radio_map_interpol_rbf} or kriging \cite{oliver1990kriging, kriging_plus_crowdsensing} are used. Although these methods are relatively fast, they are quite weak in retrieving and predicting the complex and heterogeneous spatial patterns that are usually observed in real life signals (e.g., sudden and/or highly localized transient variations in the received radio metric due to specific environmental, local or topological effects). 
Another approach consists on deterministic simulation such as Ray-Tracing tools \cite{Ray_Tracing_Paris, Laaraiedh1,Raspopoulos1, Sorour2}. Given some real field measurements and then calibrated over them, these models predict the radio propagation while simulating electromagnetic interactions with the environment. These technologies, however, need a complete description of the environment (properties of the materials of the obstacles, buildings, shape,etc.). Moreover, they are computationally complex, and in case of minor changes in the local area, these simulations should be re-run again. Recently, studies have employed machine  learning for this task by considering radio maps as images and adapting neural network models that have been proposed for image completion.   These models are based on the fully generated dataset by Ray-Tracing tools for predicting the signal propagation given the buildings mask and position of the transmitter \cite{RadioUNet}; or predicting the received power value for the Long Term Evolution (\texttt{LTE}) of the signal with use of additional information and neural networks \cite{Hayashi_variety_and_size_study, Nagao_radio_prop_prediction} with handcrafted structures. 

In this work we will focus on the received signal strength map reconstruction, where only small amount of ground truth \texttt{GPS}-tagged measurements are available preventing to use existing \texttt{NN} models with handcrafted architectures and for which Ray-Tracing models could not be applied due to the lack of information about physical properties of the environment or due to high computational complexity. Our approach is based on \textit{Neural Architecture Search} (\texttt{NAS}) \cite{nas_survey} which aims to find an optimized \texttt{NN} model for this task. We show that by employing the latter technique, it is feasible to learn model parameters while simultaneously exploring the architecture. In addition, we employ unlabeled data in conjunction with ground-truth measurements in the training phase, as well as side information that accounts for the existence of buildings, to obtain knowledge and improve the model's performance. We assess our technique using three \texttt{RSSI} Map reconstruction collections, including one we produced for the city of Grenoble in France. In the case of the latter, we thoroughly examine its properties. In particular, we show that  unlabeled data can effectively be used to find an efficient optimized \texttt{NN} model and that the side information provides valuable knowledge for learning.} The obtained model is shown to have generalization ability on base stations that were not used in the training phase. The contribution of this paper is twofold:
\begin{itemize}
    \item We propose a unified framework with the use of side information, for which we study the generalization ability of a neural network model which architecture is optimized over labeled and unlabeled data using side-information. This is an extension of the work of \cite{slnas:icann,Malkova22}.
    \item Furthermore, we provide empirical evaluation over three large-scale \texttt{RSSI} collections showing that the proposed approach is highly competitive compared to the state-of-the-art models in terms of quality metrics.
\end{itemize}

\Rep{
\section{Related State of the Art}
\label{ch1:SotA}

Classical techniques such as radial basis functions (\RBF{}) or kriging \cite{LoRa-fingerprint_outdoor} are simple and fast, but they are poor at predicting the complex and heterogeneous spatial patterns commonly observed in real-world radio signals (e.g., sudden and/or highly localized transient variations in the received signal due to specific environmental or topological effects, such as specific building shapes, presence of public furniture, ultra-navigation, etc.). Furthermore, data augmentation approaches for artificially increasing the number of measurements in radio map reconstruction issues have been developed.

The goal is to use the synthetic data created as extra data to train complex map interpolation models. However, these techniques need a highly thorough description of the physical environment and are unable to predict dynamic changes in the environment over time. A key bottleneck is their high computational complexity.

In the following, we go through some more relevant work on \RSSI{} map reconstruction, including interpolation and data-augmentation techniques, as well as machine learning approaches.

\subsection{Interpolation and data-augmentation techniques}
\label{subsec:sota_interp_tech}

Kriging or Gaussian process regression~\cite{Li_heap_overview_interp} is a prominent technique for radio map reconstruction in the wireless setting that takes into consideration the distance information between supplied measured locations while attempting to uncover their underlying 2D dependency.

Radial basis functions (\RBF{})~\cite{LoRa-fingerprint_outdoor,radio_map_interpol_rbf,rbf_interp} are another approach that simply considers the dependent on the distance between observed locations. As a result, this method is more adaptive and has been found to be more tolerant to some uncertainty \cite{rbf_kriging}. Furthermore, in order to compare the performance of the \RBF{} with different kernel functions for the map reconstruction of signal strength of LoRa radio waves,  \cite{LoRa-fingerprint_outdoor}  divided all of the points in a database of outdoor \RSSI{} measurements into training and testing subsets, with the linear kernel showing the best accuracy in both standard deviation and considered metric. The two approaches stated above (which depend on kernel techniques and underlying spatial relationships of the input measurements) need a lot of input data to provide reliable interpolation results,  making them sensitive to sparse training sets. 
These methods have consequently been considered in pair with crowdsensing, where, for example in~\cite{kriging_plus_crowdsensing}, to improve the performance of basic kriging, one calls for measuring the radio metric in new points/cells where the predicted value is still presumably imprecise. A quite similar crowdsensing method has also been applied in~\cite{SVT_RSSI_crowdsensing} after considering the problem as a matrix completion problem using singular value thresholding, where it is possible to ask for additional measurements in some specific cells where the algorithm has a low confidence in the predicted result. In our case though, we assume that we can just rely on a \RSSI{} map with 
few ground-truth initial \RSSI{} measurements.

\smallskip

Another approach considered in the context of indoor wireless localization (with map reconstruction firstly) relies on both measured field data and an a priori path loss model that accounts for the effect of walls presence and attenuation between the transmitter and the receiver~\cite{path_loss_wifi_indoor}
by using the wall matrix, which counts the number of walls along the path from the access point to the mobile location and penalize value according to that number. In some outdoor settings, training points are divided into a number of clusters of measured neighbors having specific \RSSI{} distributions, and local route loss models are applied in an effort to capture localized wireless topology effects in each cluster \cite{path_loss_rssi_outdoor_clustering}. However as parametric path loss models are usually quite imprecise, these techniques have a limited generalization capabilities and require additional impractical in-site (self-)calibration. 
A quite similar approach, except the use of additional side information, is followed in \cite{SateLoc}, where they propose an algorithm called SateLoc. Based on satellite images, it is then suggested to perform a segmentation of the areas ``crossed'' by a given radio link, depending on their type (e.g., terrain, water, forest, etc.). Then, proportionally to the size of the crossed region(s), power path loss contributions are computed according to a priori model parameters (i.e., associated with each environment type) and summed up to determine the end-to-end path loss value. 

\smallskip
One more way to build or complete radio databases 
stipulated in the context of fingerprinting based positioning  
consists in relying on deterministic simulation approach, namely Ray-Tracing tools  (e.g., \cite{Raspopoulos1, Raspopoulos2, Sorour1, Sorour2,Laaraiedh1}). 
This technique aim at predicting in-site radio propagation  (i.e., simulating electromagnetic interactions of transmitted radio waves within an environment). Once calibrated with a few real field measurements, such simulation data can relax initial metrology and deployment efforts (i.e., the number of required field measurements) to build an exploitable radio map, or even mitigate practical effects that may be harmful to positioning, such as the cross-device dispersion of radio characteristics (typicaly, between devices used for offline radio map calibration and that used for online positioning). 
Nevertheless, these tools require a very detailed description of the physical environment (e.g., shape, constituting materials and dielectric properties of obstacles, walls...). Moreover, they are notorious for requiring high and likely prohibitive  
computational complexity in real applications.
Finally, simulations must be re-run again, likely from scratch, each time minor changes are introduced in the environment, e.g. the impact of human activity (like changing  crowd  density, temporary radio link obstructions).
%
\subsection{\NN{} based models trained after data augmentation }
%

There is more and more interest in application of machine and deep learning methods to the problem of \RSSI{} map reconstruction. These approaches have shown an ability to capture unseen spatial patterns of local effects and unseen correlations. Until now, to the best of our knowledge, these algorithms were primarily trained over simulated datasets generated by data-augmentation approaches (that were mentioned above). 

\smallskip
In~\cite{RadioUNet}, given a urban environment, city geography, Tx location, and optionally pathloss measurements and car positions the authors introduce a \UNet{}-based neural network called RadioUNet in the supervised learning setting, which outputs radio path loss estimates trained on a large set of generated using the Dominant Path Model data \cite{RFB15a}.

\cite{transfer_gan_psm} propose a two-phase transfer learning with  Generative Adversarial Networks (\GAN{}), which comprises of two stages, to estimate the power spectrum maps in the underlay cognitive radio networks.
The domain projecting (\texttt{DP}) framework is used to first project the source domain onto a neighboring domain.
The target domain's entire map is then rebuilt or reconstructed using the domain completing  framework and the recovered features from the surrounding domain.

For training of the DP, fully known signal distribution maps have been used. 
In another contribution, to improve the kriging predictions the authors have used the feedforward neural network for path loss modelling ~\cite{kriging_ffnn_pl}, as conventional parametric path loss models has a small number of parameters and do not necessarily consider shadowing besides average power attenuation.

\smallskip

Apart from wireless applications, similar problems of map reconstruction also exist in other domains. In~\cite{CEDGAN} for instance, the goal is to create the full topographic maps of mountains area given sparse measurements of the altitudes values.
For this purpose, they use a \GAN{} architecture, where in the discriminator they compare pairs of the input data and the so-called ``received'' map, either generated by the generator or based on the given full true map.
In other work \cite{SST_prediction_GAN} the authors estimate the sea surface temperature with use of \texttt{GAN} architecture in the unsupervised settings but having a sequence of corrupted observations (with different clouds coverage) and known mask distribution.
%
Another close more general problem 
making extensive use of neural networks is the image inpainting problem, where one needs to recover missing pixels in a single partial image. By analogy, this kind of framework could be applied in our context too, by considering the radio map as an image, where each pixel corresponds to the \RSSI{} level for a given node location. It has been shown in \cite{slnas:icann} the interpretation of collected measurements on a map as an image with some pixels gives overall better result. But the problem is in consumed time as it does not give the generalized model and do not use the additional information about the local environment.
Usually, such image inpainting problems can be solved by minimizing a loss between true and estimated pixels, where the former are artificially and uniformly removed from the initial full image. This is however not possible and not realistic in our case, as only a few ground-truth field measurements collected on the map can be used to reconstruct the entire image. 

\smallskip

In contrast to the previous approaches, in our study, we consider practical situations where data-augmentation techniques cannot be used, mainly because of unknown environment characteristics and computational limitations, and where only a small amount of collected ground-truth measurements is available.

Finally, a few contributions aim at predicting the received power value 
based on neural networks and additional information. For instance, in \cite{Hayashi_variety_and_size_study, Nagao_radio_prop_prediction, Kazuya_Inoue_radio_prop_build}, \RSS{} values are predicted in exact points, given meta information such as the radio characteristics (e.g., transmission specifications or relationship between Rx and Tx, like horizontal/vertical angle, mechanical/electrical tilt angle, 2D/3D distance, base station antenna orientation, etc.)
and/or prior information about the buildings (e.g., height and presence). In case the latter information is missing, predictions can be made also by means of satellite images (e.g., paper \cite{Kazuya_Inoue_radio_prop_build}). In these papers though, the map reconstruction cannot be performed directly. As the prediction is realized for each point separately, it is thus time consuming. Moreover, the authors do not take into account the local signal values, but only the physical parameters and physical surroundings (similarly to standard path loss models).

\subsection{Neural Architecture Search}
\label{ch3:NAS}

The creation and selection of features in many tasks are done manually in general; this critical phase for some conventional machine learning algorithms might be time-consuming and costly. Neural Networks address this challenge by learning feature extractors in an end-to-end manner.
These feature extractors, on the other hand, rely on architectures that are still manually constructed, and with the rapid development of the field, designing an appropriate \NN{} model has become onerous in many cases.

This problem has recently been addressed by a new field of research called  (\NAS{}) \cite{nas_survey}. In a variety of applications, such as image segmentation and classification, Neural Networks with automatically found architectures have already outperformed ``conventional'' \NN{} models with hand-crafted structures.

Different types of existing methods of search are described below.
In the last few years the research on the topic of \NAS{} has been shown a huge interest in the different fields. Among various studies, there are different techniques that are based on divers methods like Reinforcement Learning  \cite{nas_rl}, Evolutionary Algorithm \cite{regu_evo} or Bayesian Optimization  \cite{nas_bo}.
Lately, recent gradient-based methods became more and more popular. For example, one of the first methods based on this technique was presented in \cite{darts} ans is called \texttt{DARTS}, which is using relaxation to, at the same time, optimize the structure of a \textit{cell}, and the weight of the operations relative to each \textit{cell}. After finding the best combinations, blocks are stacked manually to produce a neural network.
Based on \texttt{DARTS}, more complex methods have appeared such as AutoDeepLab \cite{autodeep} in which a network is optimized at 3 levels : $(i)$ the parameters of the operations, $(ii)$ the cell structure and $(iii)$ the macro-structure of the network that is stacked manually. Despite the fact that a complex representation leads to powerful architectures, this technique has some drawbacks, such as the fact that the generated architecture is single-path, which means it does not fully exploit the representation's capabilities. Moreover, as the search phase is done over a fixed network architecture, it might not be the same between different runs, thus it is complicated to use transfer learning and the impact of training from scratch can be significant. To overcome these limitations, one possible technique is to use \textit{Dynamic Routing} as proposed in \cite{li2020learning}. This approach is different from the traditional gradient based methods proposed for \NAS{} in the sense that it does not look for a specific fixed architecture but generates a dynamic path in a mesh of cells on the fly \textit{without searching} by weighting the paths during training procedure.

In the topic of signal strength map reconstruction there was no studies (for our knowledge) with application of \NAS{} to the field measurements. 

In our study, we look at how well neural networks can extract complex features and their relationships to signal strength in the local area or under similar conditions, as well as their ability to take into account additional environmental information without having access to more complex physical details. This is performed through a search for a model with an optimized architecture adapted to the task. For this we consider the genetic algorithm of architecture search as it has been shown in \cite{slnas:icann} that it outperforms the dynamic routing search.

\subsection{Semi-Supervised Learning}
\label{ch3:subsec:semi_supervised_learning}
The constitution of coherent and consistent labeled collections are often done manually. This necessitates tremendous effort, which is generally time consuming and, in some situations, unrealistic. The learning community has been looking at the concept of semi-supervised learning for discrimination and modeling tasks since the end of the 1990s, based on the observation that labeled data is expensive while unlabeled data is plentiful and contains information on the problem we are trying to solve.

\paragraph*{Framework and definitions} In this case, the labeled examples are generally assumed to be too few to obtain a good estimate of the association sought between the input space and the output space and the aim is to use unlabeled examples in order to obtain a better estimate.
For this, we will assume available a set of  labeled training examples $S = \{(x_i, y_i) \mid i = 1, \ldots, m\} \in (\mathcal X\times \mathcal{Y})^m$ supposed to be generated i.i.d. from an underlying distribution $\mathcal D$; and a set of unlabeled examples $X_u = \{x_i \mid i = m+1, \ldots, m+u\}$ that are drawn i.i.d. from the marginal distribution $\mathbb{P}(x)$.
If $X_u$ is empty, we fall back on the problem of supervised learning. If $S$ is empty, we deal with an unsupervised learning problem. During learning, semi-supervised algorithms estimate labels for unlabeled examples. We note $\tilde y$ the pseudo-label of an unlabeled example $x\in X_u$ estimated by these algorithms. The interest of semi-supervised learning arises when $u = |X_u| \gg m = |S|$ and the goal is that the knowledge one gains about the marginal distribution, $\mathbb{P}(x)$, through the unlabeled examples can provide information useful in inferring $\mathbb{P}(y\mid x)$. If this goal is not achieved, semi-supervised learning will be less efficient than supervised learning and it may even happen that the use of unlabeled data degrades the performance of the learned prediction function \cite{Zhang:00, Cozman02, Usunier11}. It is then necessary to formulate working hypotheses for taking unlabeled data into account in the supervised learning of a prediction function.

\paragraph*{Inductive vs Transductive Learning} Before presenting these hypotheses, we note that semi-supervised learning can be formulated in two different possible settings, namely \textit{transductive} and \textit{inductive} learning. The aim in inductive case is to minimize the generalization risk with respect to the distribution $\mathcal{D}$, by training a model over a finite number of training samples \cite{Vittaut02}. This setting is also the most common in semi-supervised learning.
\begin{figure}[!t]
    \centering
    \includegraphics[width=0.75\linewidth]{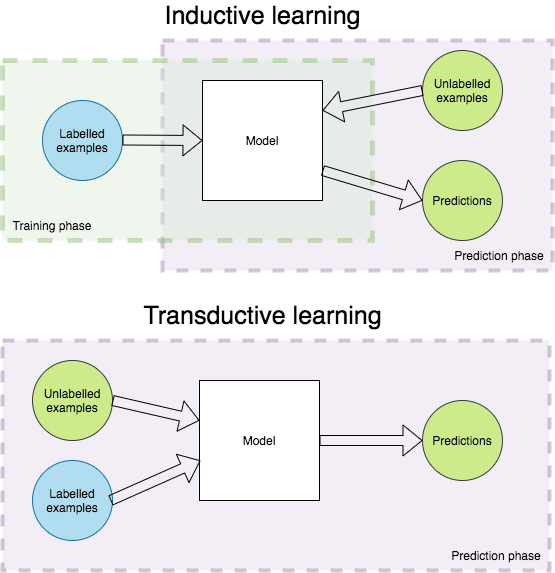}
    \caption{Inductive and transductive learning. In inductive learning there are two separate stages of training and prediction, while in transductive learning there is only one and which is prediction.}
    \label{fig:transductive_inductive}
\end{figure}

Despite this, accurate predictions for only the unlabeled cases in $X_u$ are more crucial in some applications than finding a more general rule for all existing examples drawn i.i.d. with respect to $\mathcal{D}$.
It was shown in \cite{Vapnik1982, Vapnik1998} that in this scenario, it would be preferable to ignore the more general problem and concentrate on an intermediary problem known as \textit{transductive learning} \cite{Feofanov19}. As a result, rather than looking for a general rule first (as in inductive learning), the goal of learning in this situation would be to predict the class labels of unlabeled cases by having the smallest average error (called the transductive error). These two settings are depicted in Figure \ref{fig:transductive_inductive}.

\paragraph*{Central assumptions} 

The three major hypotheses in SSL that are: \textit{smoothness} assumption, \textit{cluster} assumption and \textit{manifold assumption}. 

The basic assumption in semi-supervised learning, called the smoothness assumption states that:

\textit{$\mathcal{H}_1$: If two examples $x_1$ and $x_2$ are close in a high density region, then their class labels $y_1$ and $y_2$ should be similar.}

This assumption implies that if two points belong to the same group, then their output label is likely to be the same. If, on the other hand, they were separated by a low density region, then their outputs would be different.

Now suppose that the examples of the same class form a partition. The unlabeled data could then help find the boundary of each partition more efficiently than if only the labeled examples were used. So one way to use the unlabeled data would be to find the partitions with a mixture pattern and then assign class labels to the partitions using the labeled data they contain. The  underlying hypothesis of the latter, called the cluster assumption, can be formulated by:

\textit{$\mathcal{H}_2$: If two examples $x_1$ and $x_2$ are in the same group, then they are likely to belong to the same class~$y$.}

This hypothesis could be understood as follows: if there is a group formed by a dense set of examples, then it is unlikely that they can belong to different classes. This is not equivalent to saying that a class is formed by a single group of examples, but that it is unlikely to find two examples belonging to different classes in the same group. According to the previous continuity hypothesis, if we consider the partitions of examples as regions of high density, another formulation of the partition hypothesis is that the decision boundary passes through regions of low density. This assumption is the basis of generative and discriminative methods for semi-supervised learning. 

For high-dimensional problems, these two hypotheses may not be accurate since the search for densities is often based on a notion of distance which loses its meaning in these cases. A third hypothesis, called the manifold assumption, on which some semi-supervised models are based, then stipulates that:

\textit{$\mathcal{H}_3$: For high-dimensional problems, the examples are on locally Euclidean topological spaces (or geometric manifolds) of low dimension.}

In the following, we will present some classic models of the three families of semi-supervised methods resulting from the previous hypotheses.

There are three main families of SSL approaches that have been developed according to the above assumptions.

\paragraph*{Generative Methods}
Semi-supervised learning with generative models involves estimating the conditional density $\mathbb{P}(x \mid y, \Theta)$ using a maximum liklihood technique to estimate the parameters $\Theta$ of the model. In this case, the hidden variables associated with the labeled examples are known in advance and correspond to the class of these examples. The basic hypothesis of these models is thus the cluster assumption (Hypothesis $\mathcal{H}_2$) since, in this case, each partition of unlabeled examples corresponds to a class \cite{Seeger01}. We can thus interpret semi-supervised learning with generative models $(a)$ as a supervised classification where we have additional information on the probability density $\mathbb{P}(x)$ of the data, or $(b)$ as a partition with additional information on the class labels of a subset of examples \cite{Basu:2002,DBLP:journals/jair/MaximovAH18}. If the hypothesis generating the data is known, generative models can become very powerful \cite{Zhang:00}.

\paragraph*{Discriminant Methods}
The disadvantage of generative models is that, in the case where the distributional assumptions are no longer valid, their use will tend to deteriorate their performance compared to the case where only labeled examples are employed to learn a model \cite{Cohen:2004}. This finding has motivated many works to overcome this situation. The first works were based on the so-called directed decision technique (or self-training) proposed in the context of the adaptive processing of the signal and which consists in using the current predictions of the model for unlabeled examples in order to assign them pseudo-labels and use the pseudo-labeled examples in the training process. This process of pseudo-labeling and  learning is repeated until no more unlabeled examples are pseudo-labeled. In the case where class pseudo-labels are assigned to unlabeled examples, by thresholding the outputs of the classifier corresponding to these examples, it can be shown that the self-learning algorithm works according to the clustering assumption \cite{DBLP:journals/corr/abs-2202-12040}. 

\paragraph*{Graph-based Methods}
Generative and discriminant methods proposed in semi-supervised learning exploit the geometry of the data through density estimation techniques or based on the predictions of a learned model. The last family of semi-supervised method uses an empirical graph $G=(V,E)$ built on the labeled and unlabeled examples to express their geometry. The nodes $V=[1,\ldots,m+u]$ of this graph represent the training examples and the edges $E$ translate the similarities between the examples. These similarities are usually given by a positive symmetric matrix $\boldsymbol W=[W_{ij}]_{i,j}\in\mathbb{R}^{(m+u)\times(m+u)}$, where the weight $W_{ij}$ is non-zero if and only if te examples indices $i$ and $j$ are connected, or equivalently, if $(i,j)\in E\times E$ is an edge of the graph $G$. 

}

\section{Application to the Stated \RSS{} Map Reconstruction Problem}
\label{ch5:map_reconstruction}
Additional information could be represented in different manners, and they could be included into the algorithm in a variety of ways, such as independent channels, parallel channels inputs, directly in the learning goal, or in the ranking metric during model selection.
We adapted the proposed algorithm presented in \cite{Malkova22} for multi-channel input by combining additional context information with the data in the model's input; and we assessed the model's performance on unseen base stations that were not utilized in the learning process.


\begin{figure}[t!]
    \centering
    \includegraphics[width=\linewidth]{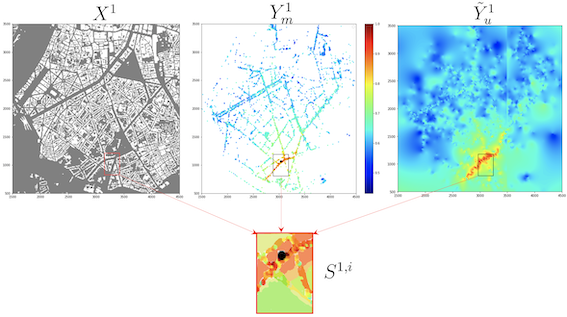}
    \caption{An example of constituting the training sets for one base station. $X^1$ corresponds to 2D node locations, buildings are shown in white. $Y_m^1$ is \RSSI{} map (true measurements); the base station is shown by a black circle,and $\tilde Y^1$ corresponds to interpolated points found by \RBF{}. Colors depict the strength of the signal from dark red (highest) to deep blue (lowest) \RSSI{} values. $S^{1,i}=S^{j,i}_\ell\cup S^{j,i}_v\cup S^{j,i}_{u\!\!\!\backslash}$ is one sub-matrix of partially labeled training data found from $Y_m^1\cup \tilde Y^1$.}
    \label{fig:image_cutting}
\end{figure}

Here, we suppose to have a small set of $n$ available base stations $(X^j)_{1\leqslant j\leqslant n}$. For each given matrix of base station $X^j; j\in\{1,\ldots,n\}$, let $Y^j\in \mathbb{R}^{H\times W}$ be its corresponding 2D matrix of signal strength values measurements, where $H\times W$ is the size (in number of elements in a grid) of the zone of interest. In  practice, we have access only to some ground truth measurements $Y^j_m$, meaning that $Y^j_{m} =~Y^j \odot M^j$, with $M^j \in \{0, 1\}^{H \times W}$ a binary mask of available measurements, and $\odot$ is the Hadamard’s product. Here we suppose sparsity meaning that the number of non-null elements in $Y^j_{m}$ is much lower than the overall size $H\times W$. For each base station $X^j$ we estimate unknown measurements $\tilde{Y}_u^j$ in $Y^j$ with a \RBF{} interpolation given $(X^j_m, Y^j_m)$, so that we have a new subset $(X^j_u, \tilde{Y}^j_u)$, where $X^j_m=X^j \odot M^j$ is the associated 2D node locations of $Y^j_{m}$ in $X^j$, and the values in $\tilde{Y}^j_u$ are initially given by \RBF{} predictions on $X^j_u$ corresponding to the  associated 2D node locations (or equivalently, the cell/pixel coordinates) with respect to the base station $X^j$ which do not have measurements. In our semi-supervised setting, the values for unknown measurements in $\tilde{Y}^j_u$ will evolve  by using the predictions of the current \NN{} model during the learning process.

We further decompose the measurements set $Y^j_{m}$ into two parts: $Y^j_{\ell}$ (for \emph{training}), $Y^j_{v}$ (for \emph{validation}), such that $Y^j_{\ell}\oplus Y^j_{v} =~Y^j_{m}$, where~$\oplus$ is the matrix addition operation. Let $X^j_{\ell}, X^j_{v}$ be the associated 2D node locations pf $Y^j_{\ell}$ and $Y^j_{v}$ in $X^j$.

In our experiments the number of base stations $n$ is small, so in order to increase the size of labeled and pseudo-labeled training samples, we cut the initial measurements maps $(Y_m^j\oplus \tilde{Y}^j_u)_{1\leqslant j\leqslant n}$ into smaller matrices which resulted into the sets $(S^{j,i})_{\substack{1\leqslant j\leqslant n\\ 1\leqslant i\leqslant m_j}}$ where the sets $S^{j,i}\subseteq Y_m^j\oplus \tilde{Y}^j_u ; \forall i\in\{1,\ldots,m_j\}$ are shifted with overlapping of the points. Each  submatrix $S^{j,i}$ is hence divided into labeled, $S^{j,i}_\ell\cup S^{j,i}_v$, and pseudo-labeled (first interpolated points using \RBF{} and then using the predictions of the current \NN{} model) $S^{j,i}_{u\!\!\!\backslash}$. To each submatrix $S^{j,i}$ corresponds a 2D location $X^{j,i} \subset X$. Figure \ref{fig:image_cutting} gives a pictorial representation of the notations.

\section{\NAS{} with Genetic Algorithm for \RSSI{} map reconstruction using side information}
\label{sec:two}
\Rep{
\UNet{} \cite{ronneberger2015unet} is one of the mostly used primary Neural Network models that can handle multiple channels and hence consider side-information as well as the \RSSI{} map on their input. As additional context (or side) information, we have considered:
\begin{itemize}
    \item information about buildings presence, which was taken from the open-source OpenStreetMap dataset \cite{osmnx_Boeing2017} -- matrix of binary 0-1 values, denoted as ``buildings map'' further (Figure \ref{fig:image_cutting} left);
    \item  amount of crossed buildings by signal from base station to each point of the map. 
    By analogy to the data representation in the indoor localisation and map reconstruction, with the amount of crossed walls by signal -- matrix of non-negative integer values, denoted as ``buildings count map'' further;
    \item information about distance from the base station. By the log-normal path loss model and corresponding \RSSI{} (\cite{book_wireless_communication}: the signal strength is proportional to $-10n\log_{10}(d)$ up to additive term, where $n$ is a path loss exponent, $d$ is a distance to base station) we can take the $-\log_{10}(distance)$ transformation to emphasize the zones closest from the base station -- matrix of continuous values, denoted as ``distance map'' further;
    \item information about the relief represented by DSM (digital surface model): terrain elevation summed with artificial features of the environment (buildings, vegetation..), see Figure \ref{fig:elevs_maps}. This information was taken from the open-source dataset\footnote{\url{https://doi.org/10.5069/G94M92HB}} provided by Japan Aerospace Exploration Agency with $30m$ accuracy -- matrix of integer values, denoted as ``elevation map'' further.
    
        \begin{figure}[!hbt]
        \centering
        \begin{subfigure}[b]{0.48\linewidth}
                \includegraphics[width=\linewidth]{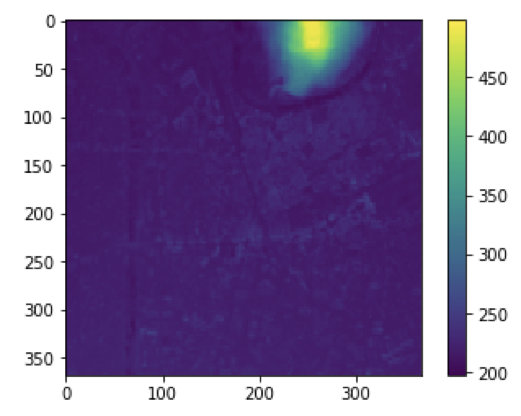}
                \caption{ Elevations map, Grenoble}
                \label{fig:elevs_Grenoble}
        \end{subfigure}
        ~ 
        \begin{subfigure}[b]{0.48\linewidth}
            \includegraphics[width=\linewidth]{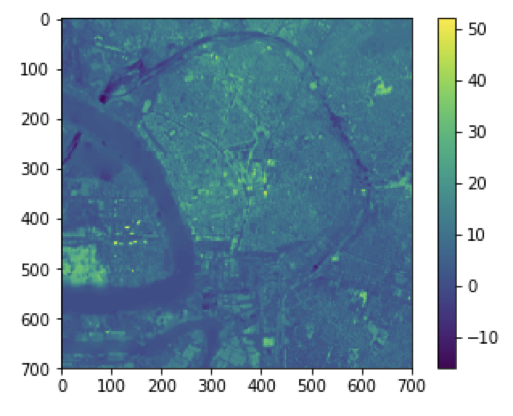}
            \caption{Elevations map, Antwerp}
            \label{fig:elevs_Antwerp}
        \end{subfigure}
        \caption{Elevation maps for two different cities}%
        \label{fig:elevs_maps}%
    \end{figure}

\end{itemize}

Our objective is to find the optimal architecture for \UNet{} using these side-information and study the generalization ability of obtained models for \RSSI{} map reconstruction.}

\NAS{} is performed with a Genetic algorithm is similar to one presented in \cite{slnas:icann} as it gave better performance result in terms of obtained accuracy. 

\smallskip

From the 
sets $(S^{j,i})_{\substack{1\leqslant j\leqslant n\\ 1\leqslant i\leqslant m_j}}$
we use an evolutionary algorithm similar to \cite{regu_evo} for searching the most efficient architecture represented as a Direct Acyclic Graph. Here, the validation sets $(S^{j,i}_v)_{\substack{1\leqslant j\leqslant n\\ 1\leqslant i\leqslant m_j}}$ 
are put aside for hyperparameter tuning. 
The edges of this DAG represent data flow with only one input for each node, which is a single operation chosen among a set of candidate operations. We consider  usual operations in the image processing field, that are a mixture of convolutional and pooling layers. We also consider three variants of 2D convolutional layers  with kernels of size
3, 5 and 7, and two types of pooling layers that compute either the average or the maximum on the filter of size 4.

\smallskip

Candidate architectures are built from randomly selected operations and the corresponding \NN{} models are trained over the set 
$(S^{j,i}_\ell)_{\substack{1\leqslant j\leqslant n\\ 1\leqslant i\leqslant m_j}}$
and its (possible) combinations with side information. The resulted architectures are then ranked according to pixel-wise error between the interpolated result of the outputs over $(S^{j,i}_v)_{\substack{1\leqslant j\leqslant n\\ 1\leqslant i\leqslant m_j}}$ and interpolated measurements given by \RBF{} interpolation by filtering out the buildings. As error functions, we have considered the Mean Absolute Error (\texttt{MAE}) or its  Normalized version (\texttt{NMAE}) where we additionally weight the pixel error according to the distance matrix value. Best ranked model is then selected for mutation and placed in the trained population. The oldest and worst in the rank are then removed to keep the population size equal to 20 models. 

\smallskip

Once the \NN{} model with the optimized parameters are found by \NAS{}, $f_{\theta}$, we consider the following two scenarios for learning its  corresponding parameters $\theta$ by minimizing  
    \begin{multline}
\label{eq:unet_loss2}
\hspace{-6mm}\mathcal{L}(f_\theta,S_{\ell}\cup S^{j,i}_{u\!\!\!\backslash})\! =\! \frac{1}{n}\sum_{j=1}^n\frac{1}{m_j}\sum_{i=1}^{m_j}\left[\frac{1}{|S^{i,j}_{\ell}|}\!\!\sum_{(x,y)\in S^{i,j}_{\ell}} \!\!\!\!\ell(y,f_{\theta}(x))  \right.\\
\left.+ \frac{1}{|S^{j,i}_{u\!\!\!\backslash}|} \sum_{(x,\tilde y)\in S^{j,i}_{u\!\!\!\backslash}} \ell(\tilde y,f_{\theta}(x))\right]
    \end{multline}

These two scenarios relate to obtaining model parameters on \textit{labeled and pseudo-labeled} measurements using just \RBF{} interpolated data (scenario 1) or predictions from a first model learnt on these data (scenario 2). The overall learning process is depicted in Algorithm \ref{algo:nas-inductive}.


\begin{algorithm}[t!]    
\caption{\SNAS$^{ind}$}
     \hspace{-5mm}\textbf{Input:} A labeled training set with given measurements: $(X^j_m,Y^j_m)_{1\leqslant j\leqslant n}$ and an unlabeled set  $(X_u^{j})_{1\leqslant j\leqslant n}$\\

\hspace{-5mm}\textbf{Init:} Using $(X^j_m,Y^j_m)_{1\leqslant j\leqslant n}$, find interpolated measurements $(\tilde{Y}^{j}_u)_{1\leqslant j\leqslant n}$ over $(X_u^{j})_{1\leqslant j\leqslant n}$ using the \RBF{} interpolation method; \\

\textbf{Step 1:} Cut the initial measurements maps $(Y_m^j\oplus \tilde{Y}^j_u)_{1\leqslant j\leqslant n}$ into smaller matrices: $(S^{j,i})_{\substack{1\leqslant j\leqslant n\\ 1\leqslant i\leqslant m_j}}$.

\textbf{Step 2:} Search the optimal \NN{} architecture 
using $(S^{j,i})_{\substack{1\leqslant j\leqslant n\\ 1\leqslant i\leqslant m_j}}$;\\

\textbf{Scenario 1:} Find the parameters $\theta^\star_1$ of the \NN{} model $f_\theta$~:\\

\begin{center}
$\theta^\star_1=\mathop{\argmin}_\theta \mathcal{L}(f_\theta,S_{\ell}\cup S^{j,i}_{u\!\!\!\backslash})$ ~~~\# (Eq. \ref{eq:unet_loss2});\\
\end{center}

\textbf{Scenario 2:} Apply $f_{\theta^\star_1}$  on unlabeled data and obtain new pseudo-labeled measurements $S^{j,i}_{u\!\!\!\backslash}$
and find the new parameters $\theta^\star_2$ of the \NN{} model $f_\theta$~:
\[
\theta^\star_2=\mathop{\argmin}_\theta \mathcal{L}(f_\theta,S_{\ell}\cup S^{j,i}_{u\!\!\!\backslash})
\] 

\hspace{-5mm}\textbf{Output:} $f_{\theta^\star_1}$ for scenario 1 or $f_{\theta^\star_2}$ for scenario 2.
    \label{algo:nas-inductive}
\end{algorithm}

\section{Evaluation Setup}
\label{ch5:settings}
\Rep{
We have considered three case studies from Paris, Antwerp (The Netherlands) and Grenoble.
In the area of data-based research and in the field of machine learning singularly, it is usually hard to find large open-source datasets made of real data. In some works however, alternatively (or as a complement) to using real data, synthetic data can be generated, for instance through deterministic simulations.

In our study, we make use of three distinct databases of outdoor \RSSI{} measurements with respect to multiple base stations. The first one was generated through a Ray-Tracing tool in the city of Paris, France. The second database, which is publicly available (See \cite{antw_dataset}), consist of real \GPS{}-tagged LoRaWAN measurements that were collected in the city of Antwerp (The Netherlands). Finally, a third database, which is also made of real \GPS{}-tagged LoRaWAN measurements, was specifically generated in the city of Grenoble (France), in the context of this study.

\subsection{Paris dataset}
\label{subsec:paris_dataset}
This first dataset is made of synthetic outdoor \RSSI{} measurements, which were simulated in a urban Long Term Evolution (\texttt{LTE}) cellular context with a ray-tracing propagation tool named VOLCANO (commercialized by SIRADEL). Those simulations were calibrated by means of side field measurements \cite{Ray_Tracing_Paris}. This kind of deterministic tool makes use of both the deployment information (typically, the relative positions of mobile nodes and base stations) and the description of the physical environment (i.e., a city layout with a faceted description of the buildings, along with their constituting materials) to predict explicitly the electromagnetic interactions of the multipath radio signal between a transmitter and a receiver. Beyond the main limitations already mentioned in section \ref{subsec:sota_interp_tech} regarding mostly computational complexity and prior information, we acknowledge a certain number of discrepancies or mismatches in comparison with the two other datasets based on real measurements. For example, in the simulated scenario, the dynamic range of observed \RSSI{} is continuous in the interval [-190, -60] dBm, while with the real measurement data, a receiver sensitivity floor of -120 dBm is imposed. Moreover, the available simulation data was already pre-aggregated into cells, thus imposing somehow the finest granularity. The overall scene is $1000m\times 1000m$, each pixel being $2m \times 2m$, thus forming a matrix of size $500 \times 500$. The area considered in these simulations is located in Paris between Champ de Mars (South-West), Faubourg Saint Germain (South), Invalides (Est), and Quai Branly / d’Orsay (North), as shown in Figure \ref{fig:bs_locations_paris}. For each pixel, the \texttt{RSS} value was simulated with respect to 6 different Base Stations. An example is given for one of these base stations in Figure \ref{fig:example_distr_Paris}. Further details regarding the considered simulation settings can be found in \cite{Ray_Tracing_Paris}.

\begin{figure}[!htb]
    \centering
    \includegraphics[width=0.7\linewidth]{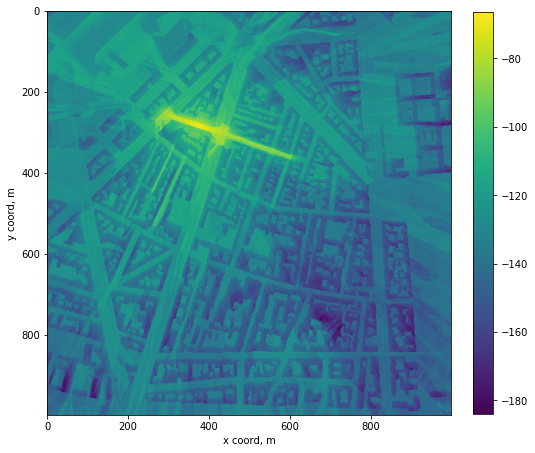}
    \caption{Example of signal strength distribution (in dBm) generated through Ray-Tracing in the Paris dataset, with respect to one particular base station roughly located in (300m, 300m).}
    \label{fig:example_distr_Paris}
\end{figure}

\begin{figure}[!htb]
    \centering
    \includegraphics[width=0.7\linewidth]{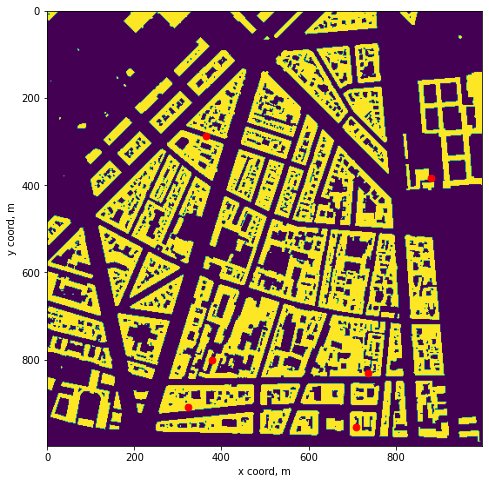}
    \caption{Buildings map and corresponding Base Stations positions (in red) for Paris dataset}
    \label{fig:bs_locations_paris}
\end{figure}

\subsection{Antwerp dataset}
\label{subsec:antwerp_dataset}

\paragraph{Measurement campaign and experimental settings}

The LoRaWAN dataset was collected in the urban are at the city centre of Antwerp from 17 November 2017 until 5 February 2018, \cite{antw_dataset}, \cite{Aernouts_thesis_antwerp_dataset}. 
The dataset consists of 123,529 LoRaWAN messages with \GPS{} coordinates on the map with \RSSI{} measurements for that location. It was collected over a network driven by Proximus (which is a nation-wide network) by twenty postal service cars equipped with The City of Things hardware. The latitude, longitude and Horizontal Dilution of Precision  information were obtained by the Firefly X1 \GPS{} receiver and then sent in a LoRaWAN message by the IM880B-L radio module in the 868 MHz band. The interval between adjacent messages was from 30s to 5 min depending on the Spreading Factor used.


The information was collected for 68 detected base stations in the initial database. We have filtered out some stations which have overall less than 10000 messages and/or which were located far from the collection zone having a flat signal. Finally we considered 9 base station -- from $BS'_1$ to $BS'_9$ (see Figure \ref{fig:bs_locations_antwerp}).
\begin{figure}[!htb]
    \centering
    \includegraphics[width=0.7\linewidth]{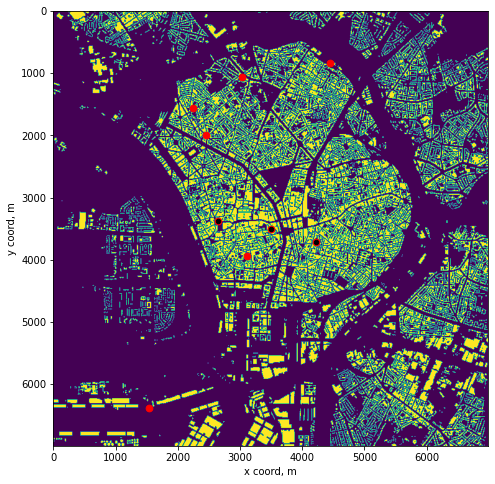}
    \caption{Buildings map and corresponding considered 9 Base Stations positions (in red) for Antwerp dataset}
    \label{fig:bs_locations_antwerp}
\end{figure}

The initial dataset 
with information about each BS or gateway (\texttt{GW}), Receiving time of the message (RX time), Spreading Factor, Horizontal Dilution of Precision, Latitude, Longitude looks as following:


\paragraph{Dataset preprocessing and analysis.}
As an example, in this part we will explain the way the dataset was processed for the future application. We aggregated the received power into cells of the size 10 meters $\times$ 10 meters ($10 m \times 10 m$) and then averaged this power and translated into signal strength. To perform this aggregation, we measured the distance from the base station location based on local East, North, Up  coordinates.

To compute the measurements density after data aggregation into cells of size $10m \times 10m$, we considered the close zone around the base station location of the size $3680m \times 3680m$. As an example, in the following Table \ref{tab:antw_dataset} there is an information about the first three considered base stations from this dataset.
\begin{table}[!t]
    \centering
   \begin{tabular}{c|c|c}
        Base station number & \multicolumn{1}{|p{1.8cm}|}{\centering Amount of measurements \\ after aggregation} & Spatial density (per $km^2$)\\
        \hline
        $BS^{\prime}_1$ & 6450 & 440\\  
        $BS^{\prime}_2$ & 5969 & 389\\ 
        $BS^{\prime}_3$ & 7118 & 525\\  
    \end{tabular}
    \caption{Amount of measurements for each base station located in the center of $368\times368$ image size after $10m \times 10m$ aggregation, Antwerp dataset. Base stations with the highest amount of measurement points around the base station location were selected.}
    \label{tab:antw_dataset}
\end{table}

We consider the zone of full city of the size $7000m \times 7000m$ which covers the positions and most of the collected measurements in the city area. We analogically did the $10m \times 10m$ aggregation. 

In the initial dataset if in the visited point on the map there was no captured signal, this point for corresponding base station was marked as -200 dBm, so in the Figure \ref{fig:Antwerp_10x10} the informative range of the signal values lies in [-120; -60] dBm, where the left boundary correspond to the sensitivity of the device.
\begin{figure}[!htb]
    \centering
    \includegraphics[width=0.6\linewidth]{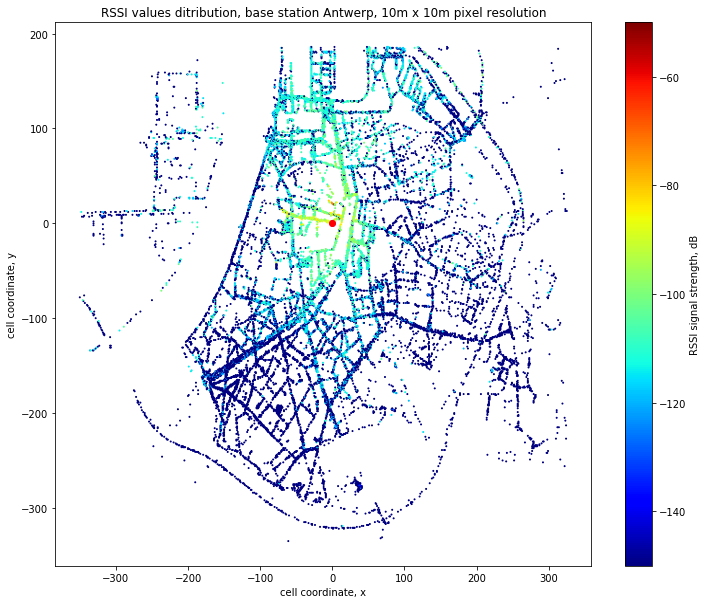}
    \caption{10x10 $m^2$ aggregation for one of the base stations, Antwerp}
    \label{fig:Antwerp_10x10}
\end{figure}

\subsection{Grenoble dataset}
\label{subsec:grenoble_dataset}

\paragraph{Measurement campaign and experimental settings.}

There was conducted (and still in progress of extension of number of measurements) an experimental campaign of collecting a LoRa measurements in the real urban environment of Grenoble city. This data is valuable for the future applications especially because of limited amount of available data for the research purposes. 
The dataset is in process of enlargement (with the installation of new base stations, extension of amount of collected measurements, etc.). 

The data is collected for several base stations installed in Grenoble and consists of several parameters such as latitude, longitude, and corresponding \RSSI{} value for each recognized base station. The collected data is similar to one described for Antwerp, but has a bit different structure, see Table \ref{tab:grenoble_dataset}.

The data was collected by different types of movements: walking of pedestrians, riding a bike or driving a car. 
Total amount of stored lines in the database collected from 13-01-2021 to 11-01-2022 is 1574588. The data was collected in the frequency band of 868 MHz in the LoRaWAN network by several users with personal tags. 
One example of stored data is shown in Table \ref{tab:grenoble_dataset}.

\begin{table*}[!htb]
    \centering
    \resizebox{\linewidth}{!}{%
    \begin{tabular}{c|c|c|c|c|c}
    $GateWay\_ID$ & $Device\_ID$ & $RSS$ & $latitude$ & $longitude$ & $time$  \\
    \hline
    7276ff002e0701e5 & 70b3d5499b4922dc & -95 & 45.199308 & 5.712627 & 2021-08-29T13:46:35\\
    7276ff002e0701f1 & 70b3d5499b4922dc & -103 & 45.199308 & 5.712627 & 2021-08-29T13:46:35\\
       \dots  & \dots & \dots & \dots & \dots & \dots \\
    \end{tabular}
    }
    \caption{Example of the dataset format for the field measurements collected in the Grenoble area, where for each connected gateway-tag pair, we report the corresponding device position, time of collection, \RSSI{} value. 
    }
    \label{tab:grenoble_dataset}
\end{table*}


To collect the data, COTS telecom grade gateways iBTS from  Kerlink manufacturer (see Figure \ref{fig:bastille_gateway}) based on the LoRaWAN technology have been used. These gateways have fine time-stamping capability, and are synchronized with the \texttt{GPS} time through a Pulse-per-Second  signal generated by \texttt{GNSS} receiver included in the gateway with an accuracy of a few nanoseconds. In LoRaWAN technology, each of the tag uplift packets can be received by more that one base station, depending on the local structure (which cause interference) and mainly path loss. To store the received information from the gateways the LoRaWAN Network Server was used, as it is shown in Table \ref{tab:grenoble_dataset} (the amount of stored metrics is bigger -- like Signal-to-Noise Ratio, Time of Arrival, uplink network parameters : frequency, DataRate, etc. but here we will focus on the data used in our study).
\begin{figure}[!b]
    \centering
        \includegraphics[width=.5\linewidth]{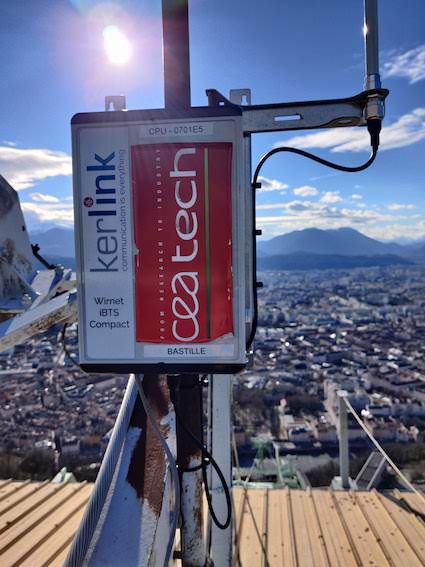}
        \caption{Gateway $BS_3$ which is installed on the roof near the cable car station at Bastille, Grenoble}
        \label{fig:bastille_gateway}
\end{figure}

During the data collection, there was different deployment characteristics of the base stations and different amount of them.  We consider the second version of the dataset (Grenoble-2) consisted of four base stations located in Grenoble and one which was far from the city region and thus was discarded because of the flatten signal in the zone with the biggest amount of measurements (the first one could be found in \cite{slnas:icann}). This version will be used in the validation of the results in the experiments section of the Section \ref{ch5:results}. Position of four considered base stations are shown in the Figure \ref{fig:BSs_positions_Grenoble}. This deployment is also interesting as one of the base stations ($BS_3$) is installed on the mountain peak higher than all the other base stations and thus quite specific to this database.

\paragraph{Dataset preprocessing and analysis.} 
First, we needed to filter out of the dataset the unreliable data resulting from errors during the data collection or from measurement artefacts. For example, the exact \GPS{} position could be significantly different from the real position (due to a lack of visibility to satellites) or, due to a specificity in the tag design, data transmission still occur while charging indoor, thus giving both the wrong \RSSI{} value and/or the wrong \GPS{} position. For instance, regarding the latter issue that is quite obvious to detect, we simply rejected all the measurements exhibiting too high \RSSI{} values and/or being static for a long time, which were most likely collected during the charging of the device. Being more precise, first we detected the base stations for which 
the received signal strength was higher than -55 dBm for the entire acquisition time sequence and then removed the corresponding measurement points collected at the same time for this device with respect to all the other base stations. 
%
However, \RSSI{} values saturating at short distances or reaching receiver sensitivity at large distances were preserved in the database, for being somehow indirectly indicative of the tag distance to the BS. 
Finally, we filtered out all the measurements for which the \GPS{} latitude was not valid (out of tolerated range).

\paragraph{Data aggregation per cell.}
Just like for the Antwr<ep dataset, after removing outliers/artefacts, we then aggregated the signal in the cells. We converted the \RSSI{} into milliWatts (as [dBm] = 10$\log_{10}$[mW]), computed its average per cell in the cells of size $10 m \times 10 m$,
and converted back the result into signal strength values.
To perform this aggregation, we measured the distance from the base station $BS_1$ location considered to be (0,0) 2D Cartesian coordinate based on local East, North, Up  coordinates.
Finally, we considered an overall area of interest of $3680m \times 3680m$ (also for the radio mapping application), which covers the entire city, while containing most of the deployed base stations, as shown in the Figure \ref{fig:BSs_positions_Grenoble}.
%
%
\begin{table*}[!htb]
    \centering
    \begin{tabular}{c|c|c}
        Base station number & \multicolumn{1}{|p{5cm}|}{\centering Amount of measurements \\ after aggregation} & \multicolumn{1}{|p{5cm}}{\centering Spatial density (per $km^2$/\\ per $10^4$ cells )} \\
        \hline
        $BS_1$ & 16577 & 1231\\ 
        $BS_2$ & 7078 &  515
    \end{tabular}
    \caption{Amount of measurements for 2 base station in the Grenoble-1 dataset (first version). Only the Base stations with the highest amount of points were selected. 
    }
    \label{tab:grenoble-1_dataset}
\end{table*}

\begin{figure}[!htb]%
    \centering
    \begin{subfigure}[b]{0.45\linewidth}
        \includegraphics[width=\linewidth]{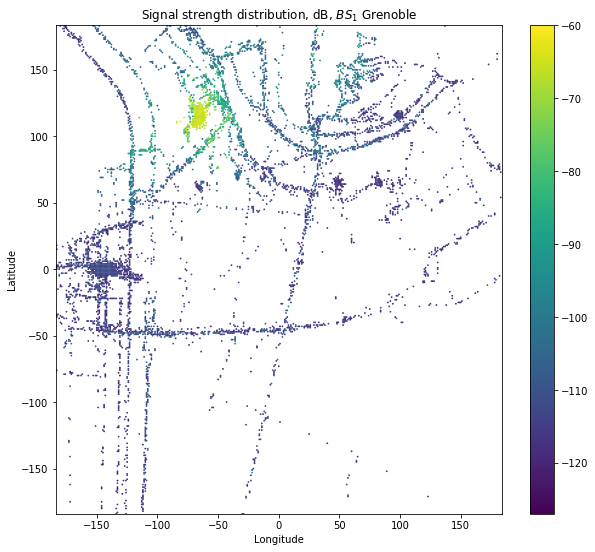}
        \caption{10x10 aggregation}
        \label{fig:10x10_signal_distr_cea_b2i}
    \end{subfigure}
    ~ 
    \begin{subfigure}[b]{0.45\linewidth}
        \includegraphics[width=\linewidth]{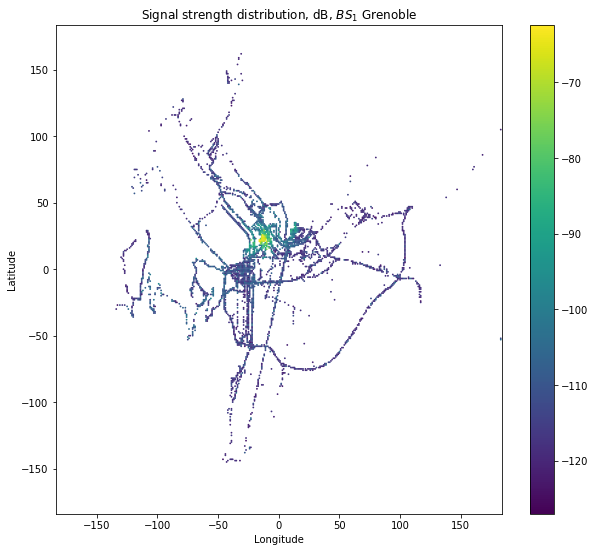}
        \caption{50x50 aggregation}
        \label{fig:50x50_signal_distr_cea_b2i}
    \end{subfigure}
    \caption{Signal distribution for different aggregation cell sizes, $BS_1$, Grenoble}%
    \label{fig:10_50_signal_distr}%
\end{figure}

To compare the measurements $std$ values per cell in different conditions, we considered two types of aggregation cell sizes: $50 m \times 50 m$ and $10 m \times 10 m$, as shown in Figure \ref{fig:10_50_signal_distr}. In case of $50 m \times 50 m$ aggregation cells, the amount of informative pixels (i.e., visited pixels with sufficient measurements) obviously reduces but at the same time, the aggregated value is more stable as a function of space (from pixels to pixels), 
while with $10 m \times 10 m$ aggregation cells, we can see significantly larger fluctuations of the average received power as a function of space but making available a larger amount of informative points for mapping. 
%
%

%
\paragraph{\RSSI{} dispersion per cell}
\label{subsec:std_distr_analysis_Gre}
%
The empirical standard deviation \emph{std} of the collected measurements per cell after data aggregation has also been calculated for the two previous cell sizes, so as to study its distribution over cells having at least 3 measurements. First, the dependence of \emph{std} on the cell size is analyzed on both Figure \ref{fig:10_50_aggreg_std} and Table  \ref{tab:std_params}. While comparing the two settings, the distribution mean looks similar, while other characteristics differ only marginally. For the $50 m \times 50 m$ cell size, it turns out that the points with high \emph{std} are mostly located closer to the base station, while for the $10 m \times 10 m$ granularity, the \emph{std} values are distributed more uniformly over the entire considered region of 368 by 368 cells.
It comes from the fact that, within typical $50 m \times 50 m$ cells close to the BS, the average \RSSI{} signal dynamic is such that the dispersion around the cell average value (i.e., the average of all the measurements collected in this cell) between the minimum and the maximum measurement values (i.e., even besides fast fading fluctuations) is naturally much larger than in the $10 m \times 10 m$ case. In other words, the fine-grain average deterministic range-dependent power decay is interpreted as extra random fluctuations in $50 m \times 50 m$ cells, due to a loose spatial grid. 
Thereby, to preserve more information about the variability of the signal while solving the problem of map reconstruction, we will consider a $10 m \times 10 m$ cell granularity in the following. 

\begin{table}[h!]
    \centering
    \begin{tabular}{c|c|c|c}
        Granularity & Mean all, dB & Median all, dB & Std, dB   \\
        \hline
        10x10 & 4.50 & 4.04 & 2.69   \\
        50x50 & 4.50 & 4.24 & 2.26 
    \end{tabular}
    \caption{Different characteristics for the distribution of the STD value per cell\\ in one street, for the points with more than 3 aggregated real \\measurements.}
    \label{tab:std_params}
\end{table}

Then, keeping a $10 m \times 10 m$ cell size, we further investigate the influence of the minimum amount of available measurement points per cell (spanning from 3 up to 30 measurements), with or without removing 10\% of the points having the largest variance of \RSSI{} measurements divided by the number of samples after in-cell data aggregation 
(See Figure \ref{fig:trsh_std_all}). This indicator indeed gives a hint on the capability to reduce fast fading dispersion through the coherent integration of in-cell instantaneous \RSSI{} measurements (i.e., variance of residual dispersion after in-cell averaging).   
After filtering out the data (Figure \ref{fig:thrsh_std_10x10_remove10}), the overall empirical distribution shape looks rather similar, even if its standard deviation is clearly decreased, as expected. This contributes typically to limit the number of cell occurrences hosting a \RSSI{} \emph{std} larger than 10dB, which are expected to very harmful to the fingerprinting process (typically, by limiting the suppression of fast-fading through averaging). 
Beyond, as a relatively limited amount of input points could be visited physically during the collection campaign within this experimental dataset, when the minimum number of points to keep the cell is too demanding, the number of exploitable cells decreases drastically while the distribution characteristics over the cells do not vary much. 
Accordingly, in terms of data preprocessing strategy, in the following (for further model parameters extraction or before applying our map interpolation algorithms), we will systematically reject 10\% of the cells with the highest \RSSI{} variance divided by number of measurements, while keeping $10 m \times 10 m$ cells with at least 3 measurements. 

\begin{figure}[!htb]%
    \centering
    \begin{subfigure}[b]{0.45\linewidth}
        \includegraphics[width=\linewidth]{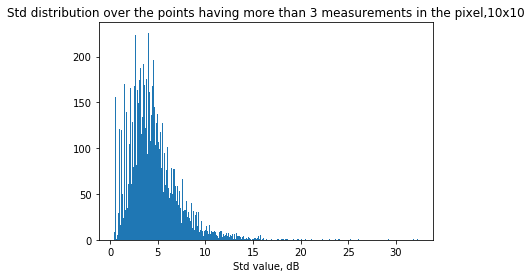}
        \caption{$10 m \times 10 m$ data aggregation cells}
        \label{fig:10x10_signal_distr_cea_std}
    \end{subfigure}
    ~ 
    \begin{subfigure}[b]{0.45\linewidth}
        \includegraphics[width=\linewidth]{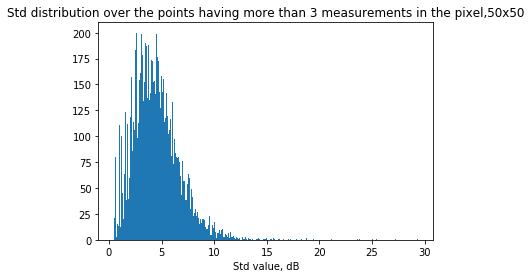}
        \caption{$50 m \times 50 m$ data aggregation cells}
        \label{fig:50x50_signal_distr_cea_std}
    \end{subfigure}
    \caption{Distribution of the standard deviation of \RSSI{} measurements per cell (over all the cells), for different aggregation cell sizes.}%
    \label{fig:10_50_aggreg_std}%
\end{figure}

\begin{figure}[!htb]%
    \centering
    \begin{subfigure}[b]{0.40\linewidth}
        \includegraphics[width=\linewidth]{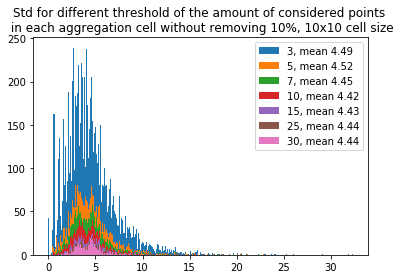}
        \caption{Without removing the points.}
        \label{fig:thrsh_std_10x10_wo_remove10}
    \end{subfigure}
    ~ 
    \begin{subfigure}[b]{0.40\linewidth}
        \includegraphics[width=\linewidth]{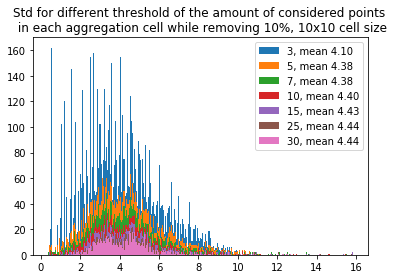}
        \caption{After removing the points.}
        \label{fig:thrsh_std_10x10_remove10}
    \end{subfigure}
    \caption{Mean value over std with different threshold over amount of points in each cell, dB, for $10m \times 10m$ cell size}%
    \label{fig:trsh_std_all}%
\end{figure}

\subparagraph{Extraction of path loss model parameters}
\label{subsec:grenoble_params}

For any tag position $\nu=(x,y)$, we consider a set of independent average received power measurements $\{P_i^{\text{dBm}}\}$, $i=1..N$ (in dBm) with respect to $N$ base stations, which are assumed to be zero-mean and normally distributed with respective variances $\{\sigma^2_i\}$, according to the classical log-normal path loss model of Eq. 
\ref{eq:pl_model}): 

\begin{equation}\label{eq:pl_model}
    P_i^{\text{dBm}}(\nu) = P_{0,i}^{\text{dBm}}- 10n\log_{10}\frac{d_i(\nu)}{d_{0,i}} + w_i, 
\end{equation}
where $d_i(\nu) = \sqrt{(x-x_i)^2+(y-y_i)^2}$ is the distance from a base station $i$ of 2D Cartesian coordinates ($x_i, y_i$) to a tag of Cartesian coordinates $\nu=(x,y)$, $P_{0,i}^{dBm}$ is the free-space average received power at the reference distance $d_{0,i}$, $i =1...N$, $w_i \sim \mathcal{N}(0,\sigma_i^2)$. For simplicity in the following, we note $ P_i^{\text{dBm}}(\nu) = P_i$, $d_{0,i}=1$ m and $P_{0,i}^{\text{dBm}}=P_{0}$, $\forall i=1..N$ and $d_i(\nu) = d_i$.

Prior to addressing more explicitly the radio map reconstruction problem (and its implications in terms of theoretical positioning performances), our goal here is first to determine empirically from real measurement data the key parameters of the path loss model introduced in Eq. \ref{eq:pl_model}, namely $n$, $P^{\text{dBm}}_0$ and $\sigma$, conditioned on propagation conditions. \textcolor{black}{The objective is indeed two-fold. First, we want to observe in practice the global trends of the received signal strength in our concrete experimental context and set-up, as a function of both the transmission range and the operating conditions (e.g., the dispersion over line-of-sight (\LoS) and non-line-of-sight (\NLoS) conditions \cite{9500705} or over serving Base Stations, the practical ranges for reaching receiver sensibility or saturation, etc.). This will indeed be helpful to qualitatively interpret the dominating factors impacting the received signal dynamics (as a function of space). One more point is also to feed an analysis based on the evaluation of theoretical positioning performance bounds, while relying on synthetic models with representative radio parameters (See next section).} 
%
%
%
%
\textcolor{black}{For this purpose, we perform Least Squares (LS) data fitting out of real field \RSSI{} measurements from the Grenoble pre-processed datasets.} 
Let us denote $\textbf{m}^{BS}$ = $(m_1^{BS}, m_2^{BS}, ..., m_K^{BS})^\intercal$
the vector of $K$ measured values of the signal strength for the corresponding base station $BS$ at distances $d_k^{BS}$, 
%
%
$\mathbf{P}(\mathbf{\eta}^{BS})=\mathbf{A}^{BS}. \mathbf{\eta}^{BS}$ 
the vector of noise-free signal strength values 
calculated for the same distances $d_k^{BS}$ based on Equation \ref{eq:pl_model}. For each base station $BS$, it hence comes:\\

\begin{equation}
    \textbf{m}^{BS} = \textbf{P}(\text{\textbf{$\eta$}$^{BS}$}) + \epsilon^{BS},
\end{equation}

where 
$\epsilon_k^{BS} \sim N(0, (\textcolor{black}{\sigma^{BS}})^2)$ are assumed to be independent and identically distributed residual noise terms (i.i.d.), 
$d_0^{BS}$ is the reference distance, $d_k^{BS}= \sqrt{(x^{BS} - x_k)^2+(y^{BS} - y_k)^2}$ is the distance from the base station of 2D Cartesian coordinates ($x^{BS},y^{BS}$) to some tag position ($x_k,y_k$). 

To compute the required model parameters, we thus need to solve the set of equations for all the measured points by minimizing the sum of squared errors, as follows:\\

\begin{equation}
    (\textbf{m}^{BS} - \textbf{A}^{BS}\mathbf{\eta}^{BS})^\intercal (\textbf{m}^{BS} - \textbf{A}^{BS}\mathbf{\eta}^{BS}) \rightarrow \min\limits_{\eta^{BS}}
\end{equation}

So the set of optimal parameters is calculated as follows :

\begin{equation}
\label{eq:intercept_pl_exponent_bs}
    \hat{\mathbf{\eta}}^{BS} = ((\textbf{A}^{BS})^\intercal\textbf{A}^{BS})^{-1}(\textbf{A}^{BS})^\intercal \textbf{m}^{BS}
\end{equation}

Retrospectively and in first approximation, residuals can be interpreted as noise in Equation \ref{eq:pl_model}, so that the standard deviation parameter $\sigma^{BS}$ is simply determined with the optimal model parameters and Equation \ref{eq:pl_model}:
\begin{equation}
\label{eq:sigma_bs}
    \hat{\sigma}^{BS} = std(\textcolor{black}{\mathbf{A}^{BS}\hat{\mathbf{\eta}}^{BS}} - \mathbf{m}^{BS})
\end{equation}
The results of this LS data fitting process for a few representative Base Stations of the Grenoble dataset 
%
%
%
are reported in Tables \ref{tab:pl_params_gre} and \ref{tab:pl_params_gre_recomp}. 
As it was mentioned above, we compared two settings with or without removing 10\% of the cells with the highest variances of measured \RSSI{} values per cell. It was done because of the possible errors or artefacts during the experimental data collection phase (e.g.,  erroneous \GPS{} position assignment due to satellite visibility conditions) or for a very small amount of collected data that were spotted to have non consistent values (e.g., due to tag failures, etc.). As we can see, removing even such a modest amount of those incriminated pathological cells can significantly impact the computation of the path loss model parameters ruling the deterministic dependency of average \RSSI{} as a function of transmission range, while the standard deviation accounting for the dispersion of \RSSI{} measurements around this average model remains unchanged.  
By the way, for all the considered Base Stations, the standard deviation value $std$ is observed to be high, which could be imputed to the fact that the underlying average path loss model is too inaccurate and can hardly account for so complex propagation phenomena. 
%
Moreover, the amount of points is not so high (maximum around 10\% of all the zone of interest), so that data fitting could be degraded by the sparseness and/or non-uniform distribution of the measurement points (with respect to the distance to the base station), which could lead to overweight the influence of some measurements in particular transmission range domains.


\begin{table}[!htb]
    \centering
    \begin{tabular}{c|c|c|c}
         BS & $\hat{n}^{BS}$ & $\hat{\sigma}^{BS}$, dB & $\hat{P}_0^{BS}$, dBm \\
         \hline
         B2I ($BS_1$) & 3.20 & 8.31 & -44.58  \\
         BCC ($BS_4$)& 2.32 & 8.73 & -55.21  \\
         Bastille ($<$500m) ($BS_3$)& 2.34 & 7.87 & -67.81 \\
         Bastille ($>$1 km) ($BS_3$)& 2.87 & 8.07 & -33.56 \\
         Bastille ($BS_2$) & 4.01 & 11.90 & -44.74  
    \end{tabular}
    \caption{Extracted path loss model parameters for some of the Base stations from Grenoble dataset, as shown on Figure \ref{fig:BSs_positions_Grenoble}, with data aggregation in $10 m \times 10 m$ cells.
    }
    \label{tab:pl_params_gre}
\end{table}

\begin{table}[!htb]
    \centering
    \begin{tabular}{c|c|c|c}
         BS & $\hat{n}^{BS}$ & $\hat{\sigma}^{BS}$, dB & $\hat{P}_0^{BS}$, dBm \\
         \hline
         B2I ($BS_1$) & 3.46 & 7.90 & -41.77  \\
         BCC ($BS_4$) & 2.59 & 8.22 & -52.18 \\
         Bastille ($<$500m) ($BS_3$)& 2.31 & 8.01 & -68.67 \\
         Bastille ($>$1 km) ($BS_3$)& 2.10 & 8.62 & -51.67 \\
         Bastille ($BS_2$) & 3.65 & 11.18 & -50.49 \\
         \hline
         $BS_1 \cup BS_2 \cup BS_4$  & 2.89 & 10.51 & -51.88
    \end{tabular}
    \caption{Recomputed path loss model parameters (after removing 10\% of the data) for the same Base stations from Grenoble dataset, as shown on Figure \ref{fig:BSs_positions_Grenoble}, with data aggregation in $10 m \times 10 m$ cells.}
    \label{tab:pl_params_gre_recomp}
\end{table}
In Table \ref{tab:pl_params_gre_recomp}, there are two sets of parameters for one of the base stations (namely "Bastille", $BS_3$), as we have identified two distinct zones in terms of topology (and hence, two propagation regimes): (i) on the hill hosting the base station, with a 
larger angular distribution of the incoming radio signals from the tags and (ii) in the rest of city where the direction of arrival of the signal at the base station does not vary much (and accordingly the receive antenna gain). 
Moreover, the position of the base station could not cover all the area around, similarly to BS ($BS_2$), being located on the Western border/part of the city, this base station naturally serves tags whose transmitted signals arrive systematically from the same side of the city (hence with a reduced span for possible angles of arrivals accordingly, which could somehow bias our extracted statistics). 
Another remark is that the path loss exponent $n$ can differ significantly from one base station to another and is usually larger in more complex environment contexts (i.e., in case denser buildings are present in the surroundings of the BS), 
%
inducing more probable \NLoS{} propagation conditions, and hence stronger shadowing effects even at short distances, which lead to lower \RSSI{} values as a function of the distance in average. But the dispersion around the fitted path loss model, as accounted here by the standard deviation $\sigma$, is also very high on its own, primarily due to a relative lack of accuracy of the single-slope path loss model, but also to the remaining effect of instantaneous received power fluctuations even after in-cell measurements averaging (e.g., caused by multipath under mobility, fast changing tag orientation during measurements collection...).

In Figure \ref{fig:model_measured_3BSs} below, we also show the overall \RSSI{} distribution over 3 base stations in town (all except the BS ``Bastille'', which again experiences a specific propagation regime due to the terrain elevation) as a function of the Tx-Rx distance, along with the superposed fitted function according to the path loss model. As expected, the dispersion accounted by the standard deviation is thus even worse here than that of previous BS-wise parameters extractions, while the other path loss parameters are close to their average values over the three considered base stations. 
%
\begin{figure}[!htb!]
    \centering
    \includegraphics[width=0.7\linewidth]{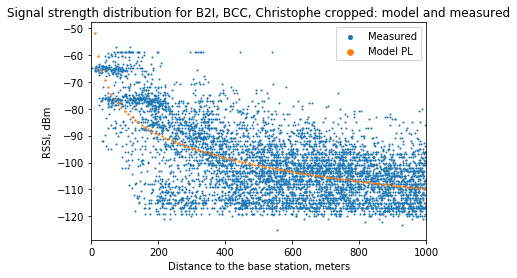}
    \caption{\RSSI{} distribution as a function of transmission range, over 3 base stations: measured values (blue) vs. model-based prediction (orange) with extracted parameters $P_0$ = -51.88dBm, $n$=2.89, $\sigma$ = 10.51 dB. 
    }
    \label{fig:model_measured_3BSs}
\end{figure}

\subparagraph{Illustration of \RSSI{} distribution along a \LoS{} street}
\label{subsec:distr_LOS}
As an illustration, we consider the longest street in \LoS{} conditions served by the Bastille Base station (Figure \ref{fig:roads_los_nlos}), where it is theoretically easier to model the behaviour of the signal distribution as a function of transmission range, due to the absence of any big obstacles on the way of the signal. Considering again the path-loss model from Eq. \ref{eq:pl_model}, the parameters have been determined through data fitting and the resulting model is confronted to the measurements, as shown in Figure \ref{fig:model_pred_bastille}. As we can see, the log-distance classical path-loss model fits fairly well the collected data in terms of average trend, but still with a very large dispersion around the expected/predicted value, indicating also that conventional parametric model-based positioning (i.e., not based on fingerprinting) would be challenging despite the \LoS{} conditions.

\begin{figure}[!t]
    \centering
    \begin{subfigure}[b]{0.4\linewidth}
        \includegraphics[width=0.95\linewidth]{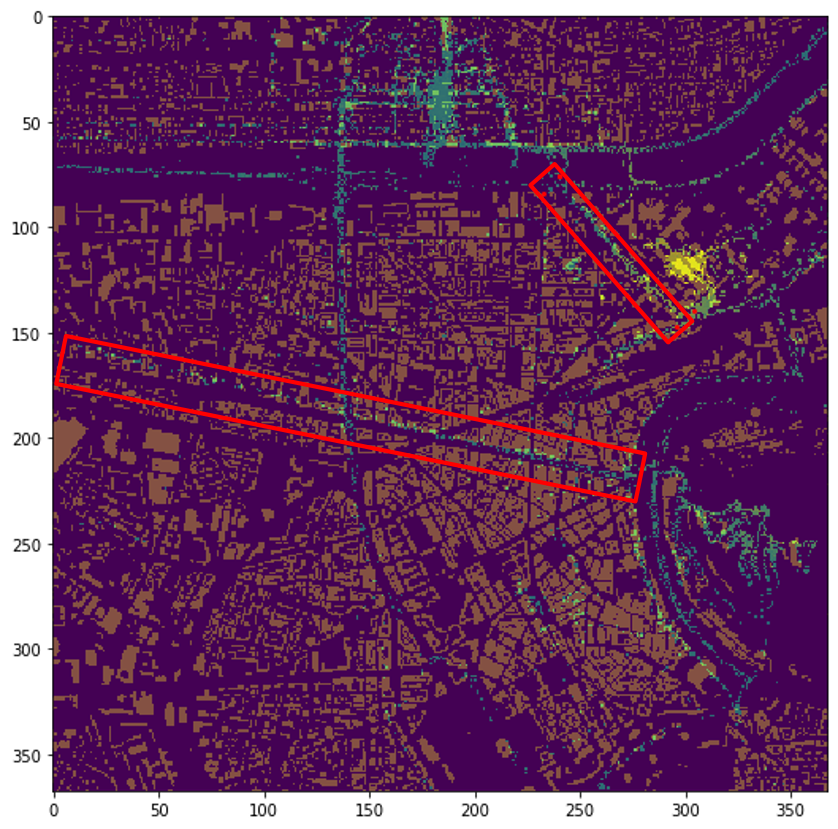}
        \caption{Example of roads in red boxes in \LoS{} with respect to at least one of the deployed BS.}
        \label{fig:roads_los_nlos}
    \end{subfigure}
    ~ 
    \begin{subfigure}[b]{0.45\linewidth}
        \includegraphics[width=\linewidth]{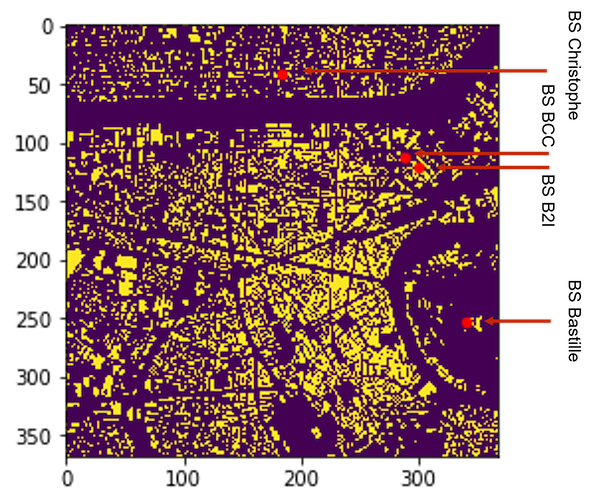}
        \caption{Positions and names of some of the base stations deployed in Grenoble.}
        \label{fig:BSs_positions_Grenoble}
    \end{subfigure}
    \caption{Part of the Grenoble map, with a selection of the deployed base station positions (right) and two canonical streets in \LoS{} with respect to the latter.}
    \label{fig:roads_los_nlos_BSs}
\end{figure}

%
\begin{figure}[!h]%
    \centering
    \begin{subfigure}[b]{0.55\linewidth}
        \includegraphics[width=\linewidth]{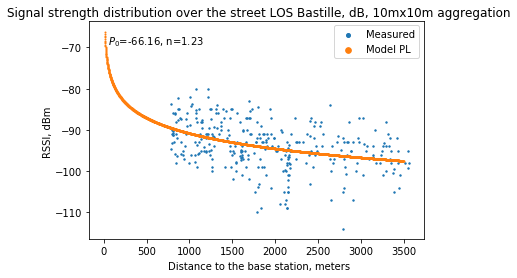}
        \caption{\RSSI{} measurements distribution and model-based 1D average prediction in \LoS{}, as a function of transmission range 
        }
        \label{fig:msm_distr_params}
    \end{subfigure}
    ~ 
    \begin{subfigure}[b]{0.35\linewidth}
        \includegraphics[width=\linewidth]{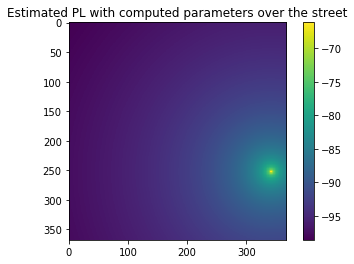}
        \caption{Model-based 2D prediction in \LoS{}, as a function of tag's position, based on extracted path loss model parameters.}
        \label{fig:ideal_distr_bast}
    \end{subfigure}
    \caption{Illustration of \RSSI{} distribution over one particular street: measured \RSSI{} values vs. model-based line-of-sight predictions.}%
    \label{fig:model_pred_bastille}%
\end{figure}
As in this \LoS{} case we can model the signal behaviour, then it is possible to compare the gradients received by this model with the possible interpolations methods and then use it as one of the quality metrics in the map reconstruction problem, the results will be shown in Section \ref{ch5:results}.

\paragraph{Datasets.}

The size of maps are 500 $\times$ 500 pixels for the generated dataset from Paris, 700 $\times$ 700 pixels for the dataset collected in Antwerp  and  368 $\times$ 368 for Grenoble. For that, as a reminder, we aggregated and averaged the power of collected measurements in cells/pixels of size 10 meters $\times$ 10 meters by the measured distance from base station location based on local ENU coordinates for Grenoble and Antwerp, while for Paris dataset the cell size is 2 meters $\times$ 2 meters. As we also consider the generalization task, the algorithm should learn from all the available base stations data simultaneously.

In our settings, we only have access to several base stations lacking several orders of magnitude in size compared to aforementioned datasets. To artificially overcome this drawback, we created submatrices of the original images by cutting them into smaller ones (we tested over 96 by 96 pixels size because of memory issues during learning of the neural network  for the storing of the model weights). We also added the flipped and mirrored images and we also did a shift in 20 pixels meaning that in our dataset there were overlapping between the images. Moreover, if the amount of pixels with measurements in the initial cutted image was high enough (more than 3\% of the presented pixels) then we masked out the randomly sampled rectangle of presented measurements similar to the cutout regularization (\cite{conv_nets_cutout_regularization}). By doing this we force the algorithm to do the reconstructions in the zones without measurements (not only locally) and be more robust to the amount of input data.

Matrices of the side information were used in the models as additional channels concatenated with measurements map.
Before feeding the data into the algorithm, all the values have been normalized between 0 and 1 in each channel separately before cutting them into smaller sizes to feed into the models. 

}
\paragraph{Evaluation of the results over held out base stations}
To evaluate the result we left one base station out of the initial set of each city to compare further the models performances with baselines, namely test Antwerp and test Grenoble. To do this, all the points were divided into two parts, namely \textit{train} and \textit{test} points for 90\% and 10\% respectively. 

\begin{figure}[h!]
    \centering
        \begin{subfigure}[b]{0.45\linewidth}
                \includegraphics[width=\linewidth]{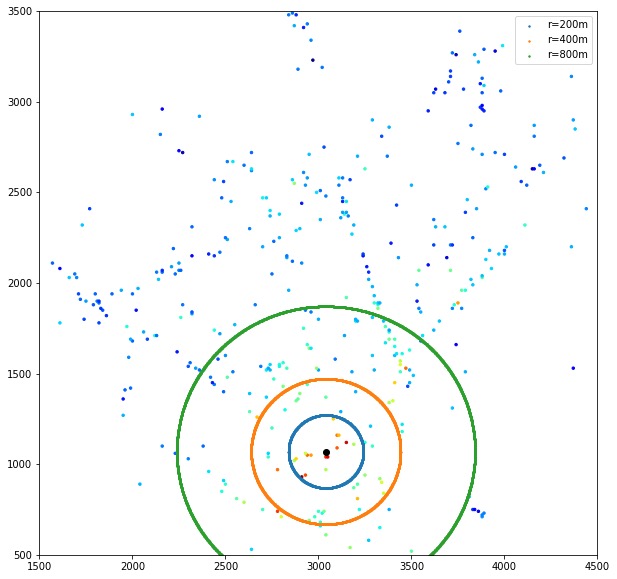}
                \caption{Test points and corresponding circular zones}
        \end{subfigure}
        ~ 
        \begin{subfigure}[b]{0.45\linewidth}
            \includegraphics[width=\linewidth]{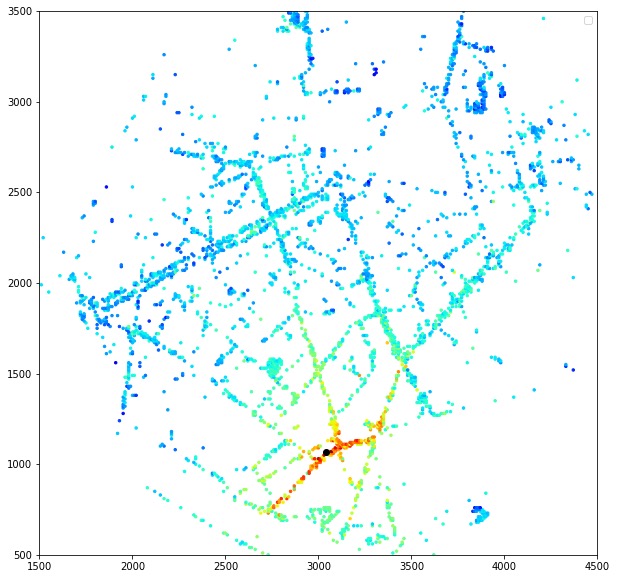}
            \caption{Train points for the same area for the same base station}
        \end{subfigure}
    \caption{Points division for the test base station, Antwerp; Base station location shown as a black point, coordinates are in meters}
    \label{fig:train_test_circles_Antwerp_val}
\end{figure}

This will be used throughout all the following sections. Moreover, to highlight the importance of the zones close to base station (as it was mentioned in the Introduction) we compare the performance of the algorithms over different considered circles around the base station location, namely 200 meters, 400 meters and 800 meters radiuses (see Figure~\ref{fig:train_test_circles_Antwerp_val}).

Information about base stations for each city, amount of all points, points, that were used in the validation/test process, and training set are given in Table~\ref{tab:summary_cities_available_points}.

We considered state-of-the-art interpolation approaches which are: Total Variation (\TV{}) in-painting by solving the optimization problem, Radial basis functions (\RBF{}) \cite{bishop2007} with linear kernel that was found the most efficient, and the $k$ Nearest Neighbours (kNN) regression algorithm.  The evolutionary algorithm in the model search phase was implemented using the \texttt{NAS-DIP} 
\cite{ho2020nas_dip} package\footnote{\url{https://github.com/Pol22/NAS_DIP}}. All experiments were run on NVIDIA GTX 1080 Ti 11GB GPU.

\section{Experimental Results}
\label{ch5:results}

In our experiments, we are primarily interested in addressing the following two questions: $(a)$ does the use of side contextual information aid in the more accurate reconstruction of \RSSI{} maps?; and; $(b)$ to what extent is the search for an optimum \NN{} design effective in the two scenarios considered (Section \ref{sec:two})?

\begin{table}[!t]
    \centering
    \resizebox{\linewidth}{!}{%
    \begin{tabular}{ll|rrrl}
    \toprule
          &      &  all points &  train points &  validation points & status \\
    city & name &         &               &                    &        \\
    \midrule
    Grenoble & $BS_1$ &    6264 &          5591 &                673 &  train \\
          & $BS_2$ &    2728 &          2448 &                280 &  train \\
          & $BS_3$ &    7266 &          6516 &                750 &  train \\
          & $BS_4$ &    6836 &          6096 &                740 &   test \\
     \hline
    Paris & $BS''_1$ - $BS''_5$ &  250000 &        7495    &               242505  &  train \\
          & $BS''_6$ &  250000 &      7495    &               242505 &   test \\
          \hline
    Antwerp & $BS'_1$ &    6060 &          5440 &                620 &  train \\
          & $BS'_2$ &    5606 &          5034 &                572 &  train \\
          & $BS'_3$ &    7548 &          6785 &                763 &  train \\
          & $BS'_4$ &    2539 &          2276 &                263 &  train \\
          & $BS'_5$ &    2957 &          2667 &                290 &  train \\
          & $BS'_6$ &    4940 &          4453 &                487 &  train \\
          & $BS'_7$ &    3154 &          2829 &                325 &  train \\
          & $BS'_8$ &    8277 &          7455 &                822 &  train \\
          & $BS'_9$ &    4335 &          3888 &                447 &   test \\
   
    \bottomrule
    \end{tabular}
    }
    \caption{Summary over the settings for different cities: total amount of available measurements, points used as an input to the models, validation (test) points that were used also in the computation of the loss (during the evaluation)}
    \label{tab:summary_cities_available_points}
\end{table}
\Rep{
Regarding the first point, we consider the following learning settings:

\begin{enumerate}
    \item given only the measurements (no side information),
    \item given both measurements and distance maps,
    \item given measurements, distances and elevation maps,
    \item given measurements, distances maps and map of amount of buildings on the way from base station to corresponding point in the map (or, in other words, buildings count).
\end{enumerate}
 
From the standpoint of application, accurate interpolation in all regions where the signal varies the most is critical. We will compare the cumulative mistakes across held-out pixels for each of the zones that are close enough to the test base station for the LoRa signal (by considering the fixed radius of 1 km).

In the following, we will present our findings using the \UNet{} model utilizing side information as a multi-channel input.

\subsection{Generalization ability of \UNet{}}
\label{ch5:subsec:unet_results}
We first study the learnability of the \UNet{} model (\cite{ronneberger2015unet}) for \RSSI{} map reconstruction without the use of unlabeled data. In order to see if there is an effect of using side information we have just considered distance maps as additional context information and considered the model with a hand-crafted classical architecture shown in Figure \ref{fig:scheme_unet_ch5} used for in-painting. 

\begin{figure}[!htb]
    \centering
    \includegraphics[width=.8\linewidth]{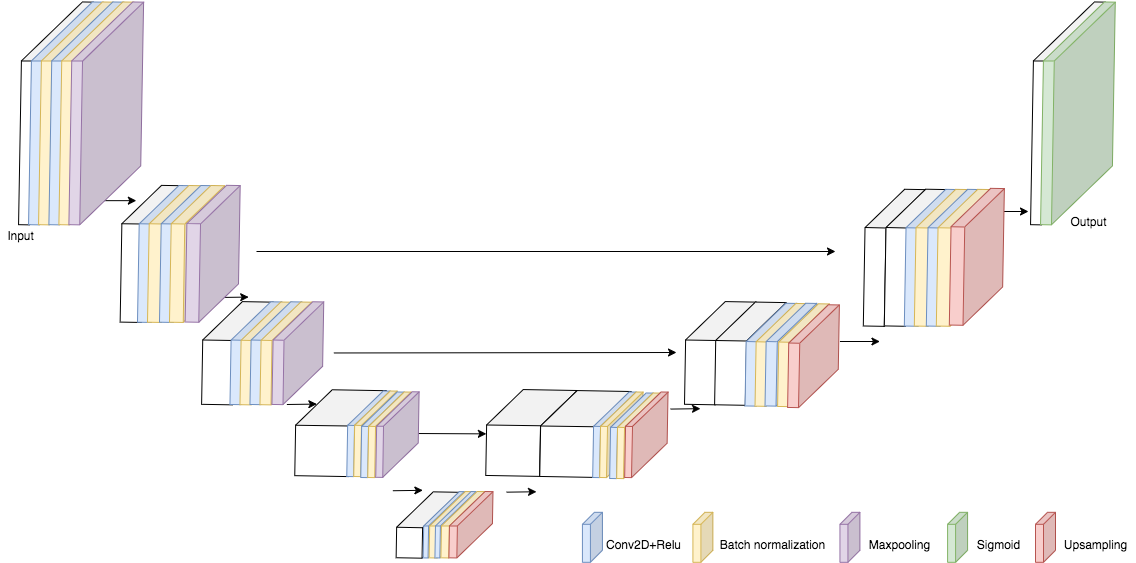}
    \caption{\UNet{} model with hand-crafted architecture proposed for in-painting.}
    \label{fig:scheme_unet_ch5}
\end{figure}

The goal of this early experiment is to validate the usage of \UNet{} for this task and investigate what effects the side information and labeled measurements have. For this we consider the simplest Paris data case where we keep only the points on the roads (as the points could be collected over the street by the vehicle drivers or pedestrians). The difference with Grenoble and Antwrep datasets is in the sampling procedure, as in reality it is very hard to obtain the collected data sampled uniformly in all the regions while in Paris dataset this is the case. All the \RSSI{} measurements also exist in Paris dataset, this allows to see the importance of labeled information in the predictions by varying the percentage of labeled measurements in the training set.

\begin{figure}[!t]
    \centering
    \includegraphics[width=.8\linewidth]{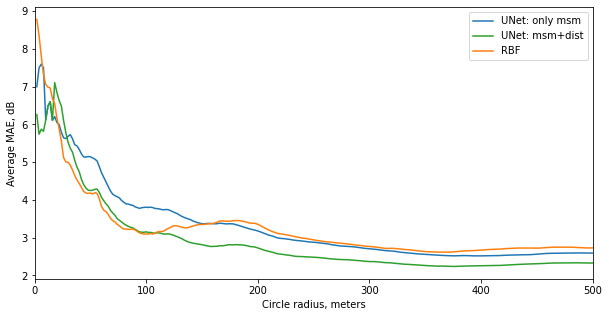}
    \caption{\texttt{MAE} for the distance from the base station for the \RBF{} and fixed \UNet{} outputs over Paris dataset with validation set compliment to the train one, measurements and distances or only measurements input}
    \label{fig:paris_dist_msm}
\end{figure}

Figure \ref{fig:paris_dist_msm} depicts the evolution of \texttt{MAE} in measurements with respect to the distance to the test base station getting lower in comparison with the \RBF{} interpolation, as well as the overall error becomes smaller ($BS''_6$ - table \ref{tab:summary_cities_available_points}), of \RBF{}, \UNet{} using only measurements (\UNet{} only msm) and \UNet{} using measurements and distance maps (\UNet{} msm+dist). From these results, it comes that \RBF{} outperforms \UNet{} using only measurements in a circle zone of less than 150 m radius around the base station. However, when distance maps are added to the model’s second channel, the situation is reversed. With the inclusion of side-information, we see that \UNet{} performs around 1 point better in \texttt{MAE} than \RBF{}. This situation is illustrated on the map reconstruction ability of both models around the test base station $BS''_6$ in Figure \ref{fig:Paris_reconstr_UNet_RBF}. As can be observed, the projected signal levels are more discernible on the roads, which are actually the zones of interest where the signal is sought, as predicted by the \UNet{} model.
\begin{figure}[!htb]
    \centering
    \begin{tabular}{cc}
    \includegraphics[width=0.45\linewidth]{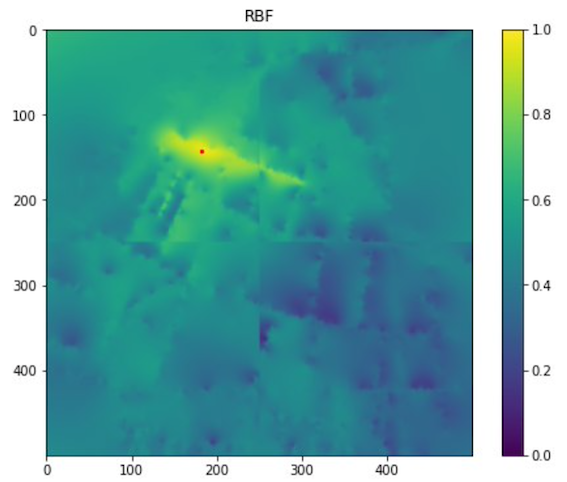}
    &
    \includegraphics[width=0.45\linewidth]{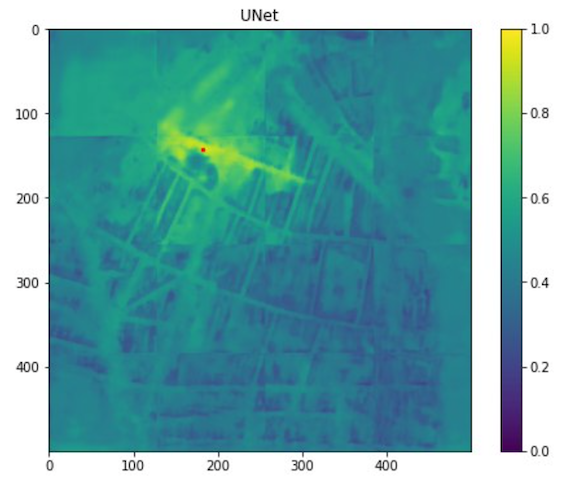}
	\end{tabular}    
    \caption{Map reconstruction of \RBF{} (left) and \UNet{} (right) over the test base station $BS''_6$ in Paris shown by a red dot.}
    \label{fig:Paris_reconstr_UNet_RBF}
\end{figure}
As a result, these findings show that the \UNet{} model can effectively account for side-information. We examined the \RBF{} and \UNet{} models for the influence of labeled measurements on the predictions by altering the percentage of labeled data utilized by \RBF{} for discovering the interpolation and by the \UNet{} model for learning the parameters. With regard to this proportion, Figure \ref{fig:paris_mae_vs_perc_of_val}  displays the average \texttt{MAE} 200 meters (left) and 400 meters (right) away from the test base station. The test error of the \RBF{} model on unseen test data remains constant as the quantity of labeled training data increases, but the test error of the \UNet{} model decreases as this number increases.

\begin{figure}[!h]
    \centering
    \begin{subfigure}[b]{0.45\linewidth}
            \includegraphics[width=\linewidth]{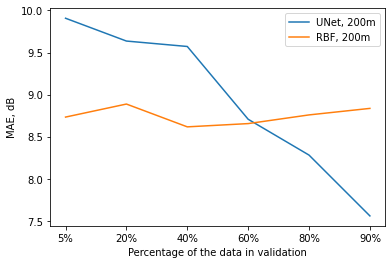}
            \caption{ 200 meters from the base station}
            \label{fig:200m_paris}
    \end{subfigure}
        ~ 
    \begin{subfigure}[b]{0.45\linewidth}
            \includegraphics[width=\linewidth]{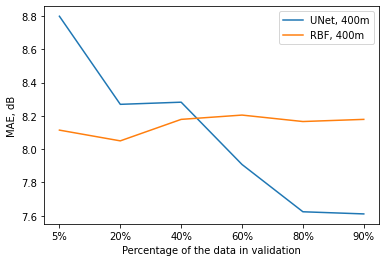}
            \caption{ 400 meters from the base station}
            \label{fig:400m_paris}
    \end{subfigure}
    \caption{\texttt{MAE} with respect to different percentage of labeled data in the training set 200 meters (left) and 400 meters (right) away from the test base station.}
    \label{fig:paris_mae_vs_perc_of_val}
\end{figure}

}
\subsection{The use of unlabeled data by taking into account side information with \NAS{}}
\label{ch5:subsec:nas_results}
\begin{figure}[!b]
    \centering
    \begin{tabular}{cc}
    \includegraphics[width=.5\linewidth]{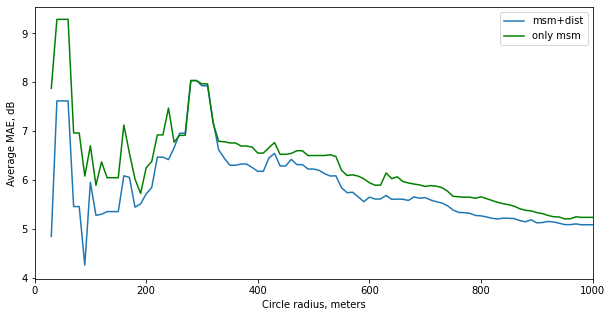} & 
    \includegraphics[width=.5\linewidth]{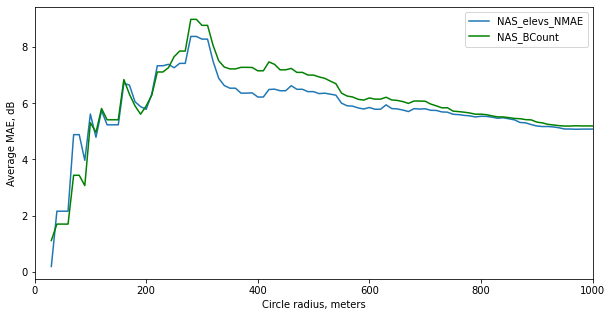}
	\end{tabular}    
    \caption{Cumulative \texttt{MAE} distribution of $f_{\theta^\star_1}$ (scenario 1 of Algorithm \ref{algo:nas-inductive}) according to the distance to the test base station for the city of Antwerp $BS'_9$; with measurements and measurements with distance maps (left) and measurements, distance maps, and building counts or elevation (right).}
    \label{fig:nas_msm_vs_msm_dist_1}
\end{figure}
\begin{table*}[!t]
    \begin{minipage}{.45\linewidth}
        \centering
        \begin{tabular}{ c | c| cpppp  }
            \hline
             Model & \texttt{MAE}, dB, 200m & \texttt{MAE}, dB, 400m \\
             \hline
             \RBF{}\cite{bishop2007} &  8.34$^\downarrow$  &  7.04$^\downarrow$   \\
             kNN &  7.98$^\downarrow$  &  7.08$^\downarrow$     \\
             TV  &  7.50$^\downarrow$  &  6.97$^\downarrow$     \\
             \UNet{} \cite{ronneberger2015unet}&  6.37$^\downarrow$  &  6.81$^\downarrow$    \\
             DIP \cite{dip}&  6.55$^\downarrow$  &  6.63$^\downarrow$    \\
             $f_{\theta^\star_1}$ &  \textbf{5.88}  &  \textbf{6.37}     \\
             \hline
        \end{tabular}
    \end{minipage}
    \begin{minipage}{.45\linewidth}
    \centering\Rep{
    \begin{tabular}{ c | c| cpppp  }
        \hline
         Model & \texttt{MAE}, dB, 200m & \texttt{MAE}, dB, 400m \\
         \hline
         \RBF{} \cite{bishop2007} &  4.03$^\downarrow$  &  5.29$^\downarrow$   \\
         kNN &  3.84$^\downarrow$  &  4.92$^\downarrow$     \\
         TV  &  4.53$^\downarrow$  &  5.91$^\downarrow$     \\
         DIP \cite{dip} &  4.64$^\downarrow$  &  5.50$^\downarrow$    \\
         $f_{\theta^\star_1}$ &  \textbf{3.40}  &  \textbf{4.32}     \\
         \hline
    \end{tabular}}
    \end{minipage}
    \caption{Comparisons between baselines in terms of \texttt{MAE} with respect to the two distances to Antwerp' test base station $BS'_9$ (left), and Grenoble test base station $BS_4$ (right). Best results are shown in bold.}
    \label{tab:MAE_results}
\end{table*}
We now expand our research to real-world data sets from Grenoble and Antwerp, taking into account more side-information and investigating the impact of neural architecture search on the creation of a better \NN{} model.  As in this case, the labeled training measurements are scarce we examine the usage of unlabeled data in addition to the labeled measurements as described in the previous section. 

We begin by envisaging scenario 1 of Algorithm \ref{algo:nas-inductive} (Section \ref{sec:two}) and investigating the impact of side information on the performance of the optimized \NN{} model discovered by \NAS{}.

\begin{figure}[!b]
    \centering
        \includegraphics[width=.8\linewidth]{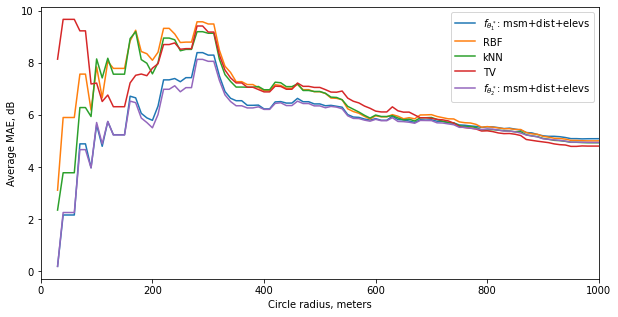}
    \caption{Average \texttt{MAE} in dB of all models with respect to the distance to the test base Station $BS'_9$ for the city of Antwerp.}
    \label{fig:mae_vs_dist_outp_all}
\end{figure}
Figure~\ref{fig:nas_msm_vs_msm_dist_1}  shows the evolution of \texttt{MAE} inside a circular zone of varied radius for $f_{\theta^\star_1}$ trained just on measurements and measurements with distance maps (left) and $f_{\theta^\star_1}$ trained on measurements, distance maps, and building counts or elevation (right) on the test base stations of Antwerp $BS'_9$. The use of distance maps to supplement measurements improves predictions, which is consistent with our earlier findings. When the third side-information is included, such as height or building counts, we find that the elevation yields better signal estimations than the latter. This is understandable because signal transmission can be severely slowed by building heights.

As a best model obtained by Algorithm~\ref{algo:nas-inductive}, scenario~1  we consider the case with three input channels: measurements, distances and elevations and present comparative results with other baselines in Table~\ref{tab:MAE_results}. The lowest errors are shown in boldface. The symbol $^\downarrow$ denotes that the error is significantly greater than the best result using the Wilcoxon rank sum test with a p-value threshold of 0.01. According to these findings, $f_{\theta^\star_1}$ outperforms other state-of-the-art models as well as the \UNet{} model with a handcrafted architecture. These results suggest that the search of an optimal \NN{} model with side-information has strong generalization ability for \RSSI{} map reconstruction.

Figure \ref{fig:mae_vs_dist_outp_all} depicts the average \texttt{MAE} in dB of all models as well as the \NN{} model $f_{\theta^\star_2}$ corresponding to scenario 2 of Algorithm \ref{algo:nas-inductive}, with respect to the distance to the test base Station $BS'_9$ for the city of Antwerp. For distances between 200 and 400 meters, $f_{\theta^\star_2}$ consistently outperforms in terms of \texttt{MAE}. As in paper \cite{slnas:icann}, these findings imply that self-training constitutes a promising future direction for \RSSI{} map reconstruction.

\Rep{

Figure \ref{fig:Grenoble_MAE_vs_dist} presents the average \texttt{MAE} in dB of all models with respect to the distance to the test base Station $BS_4$ for the city of Grenoble. These results are consistent with those obtained over the city of Antwerp.

\begin{figure}[h!]
    \centering
        \begin{subfigure}[b]{0.65\linewidth}
            \includegraphics[width=\linewidth]{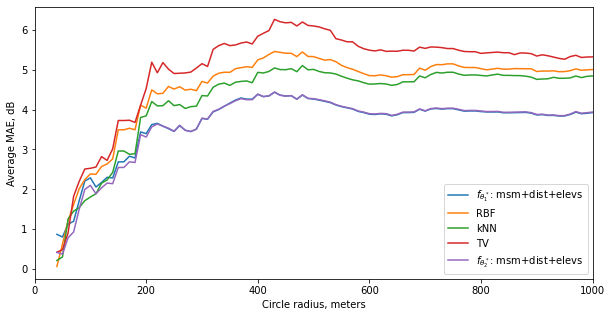}
            \caption{Comparison between different approaches}
            \label{fig:gren_all}
        \end{subfigure}
        ~ 
        \begin{subfigure}[b]{0.3\linewidth}
            \includegraphics[width=\linewidth]{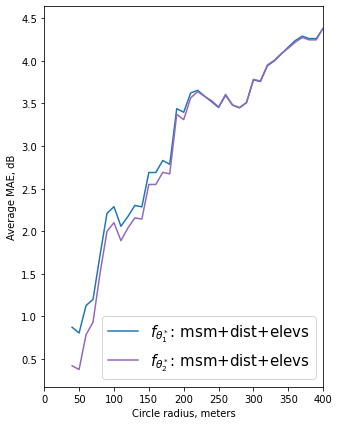}
            \caption{Zoom distance \textless 400m}
            \label{fig:gren_pseudo_lab_init}
        \end{subfigure}
    \caption{Average \texttt{MAE} in dB of all models with respect to the distance to the test base Station $BS_4$ for the city of Grenoble.}
    \label{fig:Grenoble_MAE_vs_dist}
\end{figure}

\begin{figure*}[!t]
    \centering
\begin{tabular}{cc}
                \includegraphics[width=.45\linewidth]{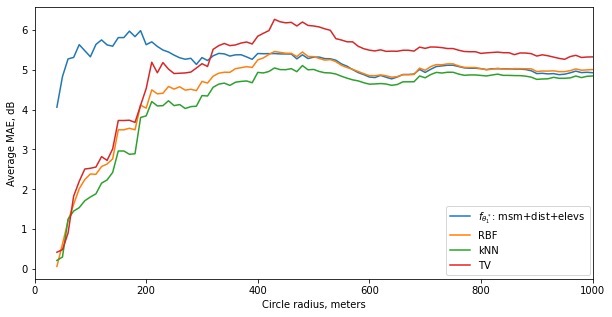}&
                \includegraphics[width=.45\linewidth]{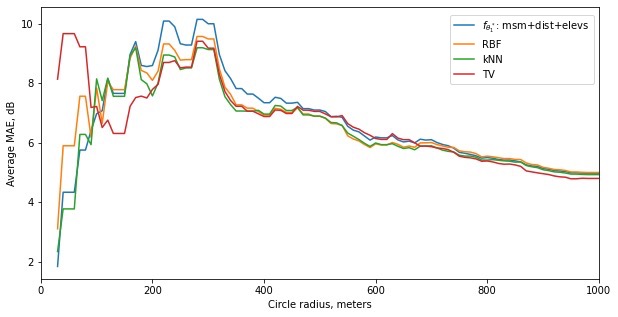}
\end{tabular}
    \caption{\texttt{MAE} over the distance to the base station evaluated over unseen base stations for Grenoble (left) and Antwerp (right), $f_{\theta^\star_1}$ is trained over mixed dataset Grenoble+Antwerp.}
    \label{fig:Antw_Gre_MAE_vs_dist_mixed_model}
\end{figure*}

The general conclusion that we can draw is that knowing about local patterns (even if from different locations/distributions/base stations) allows us to use this information in signal strength map reconstruction for application to unseen measurements from different base stations, demonstrating the ability to generalize output in the same area.    
    
In order to  get a finer granularity look at the estimations of the suggested technique, $f_{\theta^\star_2}$, Figure \ref{fig:Antw_Gre_test_points_circle_heatmap} depicts the errors heatmaps on circular zones of radius 200m and 400m surrounding the test base stations for Antwerp and Grenoble.  Each point reflects the difference between the real and predicted signal values. For both cities, we  notice that there is 
\begin{itemize}
\item  an overestimation of the signal (higher predicted values than the true ones) within the zone of radius less than 200 meters where the values of the true signal are high. In absolute value, the average \texttt{MAE} in dB are respectively 3.6 
for Grenoble and 6.3
for Antwerp.
\item an underestimation of the signal (lower predicted values than the true ones) within the zones of radius between 200 and 400 meters  where the values of the true signal are low. In absolute value, the average \texttt{MAE} in dB are respectively 4.9
for Grenoble and 6.2 
for Antwerp. 
\end{itemize}
   \begin{figure}[!h]
    \centering
\includegraphics[width=\linewidth]{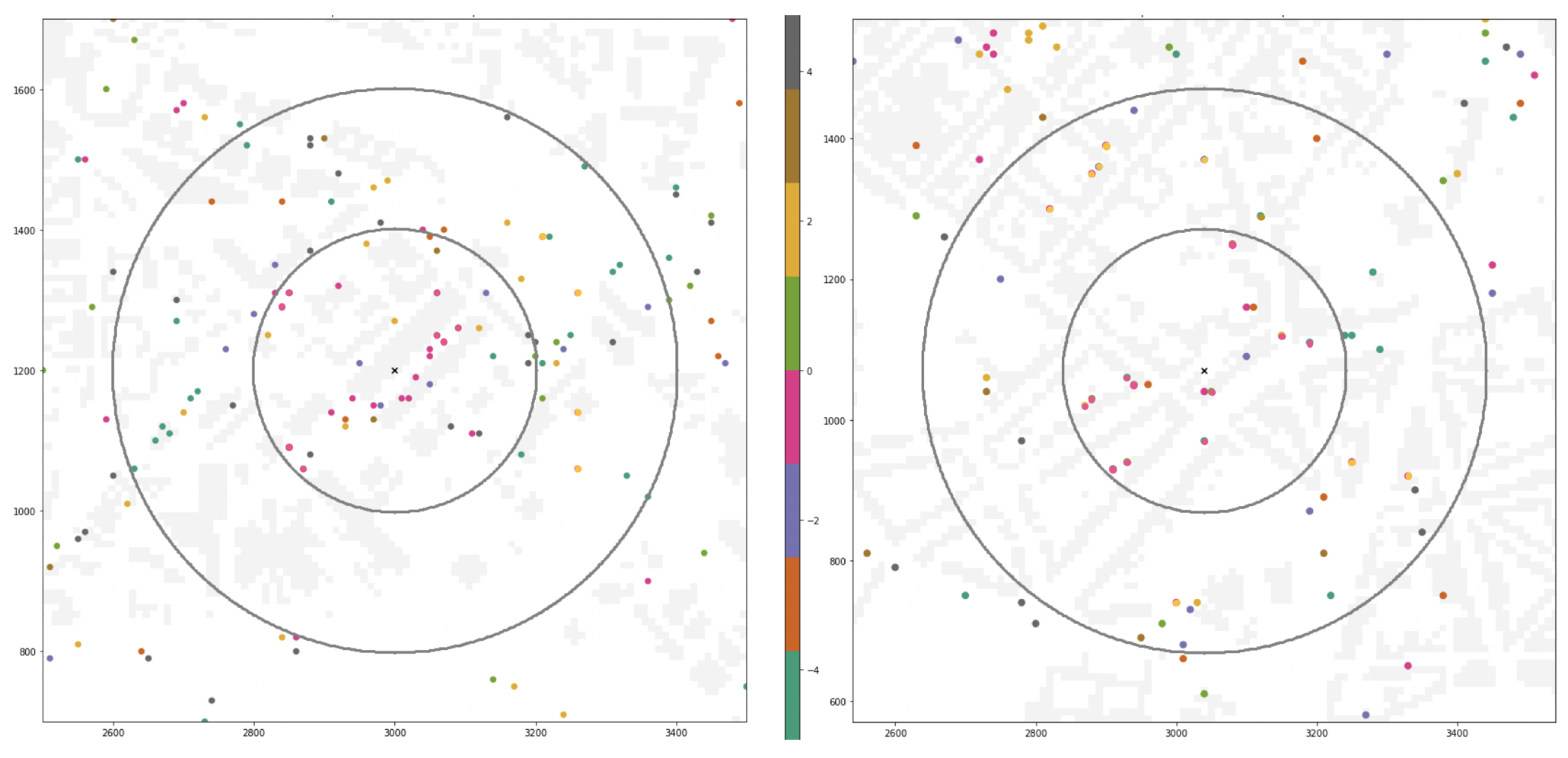}
    \caption{Heatmap of the errors, dB, between true and predicted values of, $f_{\theta^\star_2}$, over test base stations in Grenoble (left) and Antwerp (right).}
    \label{fig:Antw_Gre_test_points_circle_heatmap}
\end{figure}

To better understand the aforementioned results, we provide the empirical cumulative distribution function  of different techniques in a 200-meter zone around the test base stations in Grenoble (Figure \ref{fig:200m_antw_gre}, left) and Antwerp (Figure \ref{fig:200m_antw_gre} right). From these results, it comes that the probabilities of having less absolute dB error is higher for both $f_{\theta^\star_1}$ and $f_{\theta^\star_2}$ than the other approaches.

\begin{figure}[!h]
        \centering
        \begin{tabular}{cc}
            \includegraphics[width=0.45\linewidth]{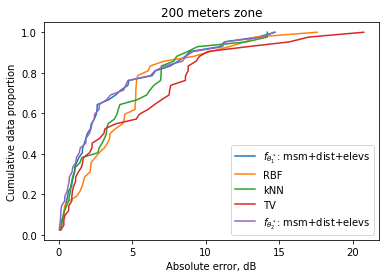}
            &
            \includegraphics[width=0.45\linewidth]{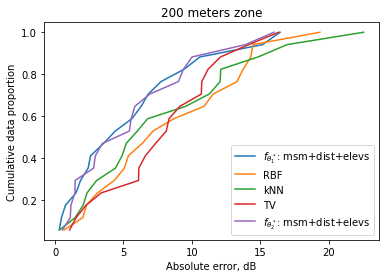}\\
           (a) & (b)
        \end{tabular}
        \caption{Empirical cumulative distribution function of different techniques in a 200-meter zone around the test base stations in Grenoble (left) and Antwerp (right).}
        \label{fig:200m_antw_gre}
    \end{figure}
The primary takeaway from these findings is that searching a Neural Network model with generalization capabilities might be useful for \RSSI{} map reconstruction. To further investigate in this direction, we considered Scenario 1 of Algorithm \ref{algo:nas-inductive} in which the training points of both cities are combined, with the goal of evaluating the model's ability to produce predictions for one of the cities. The average \texttt{MAE} in db with respect to the distance to the base stations for different approaches are shown in Figure \ref{fig:Antw_Gre_MAE_vs_dist_mixed_model}.

According to these findings, the inclusion of signal data from another city disrupts the search for an efficient \NN{} model and learning its parameters. This is most likely owing to the fact that the data distributions in these cities differ, and it would be interesting to study over alignment strategies, such as those proposed for domain adaptation \cite{DBLP:conf/nips/KumarSWKFFW18}, in order to narrow the gap between these distributions in future work.
}

\section{Conclusion}
In this paper we studied the importance of the use of additional side information for the search of an optimized \NN{} architecture for \RSSI{} map reconstruction over three different datasets. We have shown that the addition of distance and elevation of buildings to the measurements allow to significantly reduce the mean absolute error in dB of the obtained \NN{} model with an optimized found architecture. Our proposed approach tends to outperform agnostic techniques especially in the close zone near to the test base stations. We have also shown that our \NN{} based approach has good generalization ability. However, in situations where there exists a distribution shift between two maps, the prediction confidence given by the training model may be highly biased towards, and thus may be not reliable.
In reality, a significant difference between two different \RSSI{} maps could lead to complete degradation of the model's performance due to the large error in pseudo-labels. In practice, approaches like confidence regularization \cite{Zou:2019} may reduce the number of wrong pseudo-labels, but theoretically, studying the semi-supervised learning under a distribution shift is an important direction for future work.

\bibliographystyle{plain}
\bibliography{refs}

\end{document}